\documentclass[11pt]{article} 
\usepackage{amsmath, amssymb, graphics,hyperref}

\title{Formulas for SU(3) Matrices}  
\author{{\it Richard Shurtleff~}\thanks{affiliation and mailing 
address: Department of Applied Mathematics and Sciences, 
Wentworth Institute of Technology, 550 Huntington Avenue, 
Boston, MA, USA, ZIP 02115, telephone number: (617) 989-4338, 
e-mail addresses: shurtleffr@wit.edu, momentummatrix@yahoo.com}} 
\begin{document} 
          
\maketitle 

\begin{abstract} 
Formulas are developed for the eight basis matrices $\{T^{\pm},T^{3},V^{\pm},U^{\pm},U^{3}\}$ of the finite dimensional $(p,q)$-irreducible representation of SU(3). Two computer programs, one in an interpretive language and one in a compiled language, are included. Given $p$ and $q,$ each calculates the eight basis matrices.

\vspace{0.2cm}
 
\vspace{0.2cm}


\end{abstract}


\section{Introduction} \label{intro}

When working with a group, it is sometimes convenient to have numerical matrix representations. Explicit formulas for basis matrices of SU(3) have been available for decades. \cite{BandB} However the earlier versions anyway are not written in the language of currently popular elementary particle physics texts.\cite{GandG} Given the volume of literature since SU(3) became important in particle physics it is likely that formulas are available in the popular language and have been coded into computer programs. However, the author is unaware of any. This article includes a computer program and uses the notation and quantities found in popular textbooks. 

We start from the properties of the states in a $(p,q)$ irreducible representation (irrep) and obtain basis matrices that satisfy the well-known commutation relations of the algebra su(3). No attempt to show the equivalence with the earlier work of others is attempted.   

For SU(3), there are eight basis matrices. Three of the eight, popularly called $T^{\pm}$ and $T^{3},$ represent an SU(2) subgroup and can be written with standard spin matrices once the T-spins of the states have been determined. For each state, one can also determine by conventional methods the `3-component' of its U-spin, another SU(2) subgroup. The 3-components form a fourth, diagonal matrix, $U^{3}$. Of the final four matrices, two are the transposes of the other two, leaving two, $U^{+}$ and $V^{+},$ to find by various means.

The matrices $U^{+}$ and $V^{+}$ were obtained, in part, by solving the commutation relations and hermiticity requirements. Less rigorously, some expressions were developed by trial and error using a numerical catalog of matrices  for small $(p,q)$ found by computer. 

The matrices here are written for basis states distinguished by $\{p,q,s,\sigma,u^{3}\}$ where $(p,q)$ labels the irrep, $s^2$ is the square of the T-spin, $\sigma$ is one component (the `3-component') of the T-spin, and $u^{3}$ is one component (another `3-component') of the U-spin. States first in line are those with the lowest T-spin $s$ then by the 3-component of T-spin, largest $\sigma$ first. A set of $2s+1$ states with the same T-spin $s$ but lower 3-components of U-spin precedes similar sets with larger U-spin 3-component $u^{3}.$ 

As a practical matter, those using the formulas are advised to check for the proper trace and hermiticity of the matrices, ensure the examples satisfy the commutation relations and compare the Casimir invariant with well-known formulas. Matrices obtained with the formulas developed herein have been checked by computer for all $(p,q)$ irreps with $1 \leq$ $p + q \leq$ $10.$

The eight basis T,U,V matrices of a $(p,q)$-irrep can be found with a Mathematica \cite{Wolfram} notebook that is presented in Sec. \ref{mma} and with a Fortran 90 program \cite{fortran90} that is presented in Sec. \ref{fortran}. To run, export the programs to plain text. The Mathematica program requires inserting carriage returns where needed. Links to ready-to-run files are provided for convenience in Ref. \cite{notebook} and  Ref. \cite{f90}.

\section{The Fundamental Rep} \label{10}

Unitary $3\times 3$ matrices with determinant equal to one form a group under matrix multiplication; the product of two unitary matrices is unitary and the hermitian conjugate of a unitary matrix is its inverse and is unitary. 

A unitary $3\times 3$ matrix $D(\theta)$ can be written as a function of eight real parameters $\theta_{i},$ $i \in$ $\{1,2,...,8\}$ with eight generators $F^{i},$
\begin{equation} \label{expiaF} D(\theta) \equiv \exp{i \theta_{i} F^{i}}   \quad .
\end{equation}
It follows that the generators $F^{i}$ are traceless $3\times 3$ hermitian  matrices.

A well-known set of $F^{i}$s is \cite{GandG}
\begin{equation} \label{Gell-Mann} F^{1} = \frac{1}{2} \begin{pmatrix}0&&1&&0 \cr 1&&0&&0 \cr 0&&0&&0 \end{pmatrix} \quad F^{2} = \frac{1}{2}\begin{pmatrix}0&&-i&&0\cr i&&0&&0\cr0&&0&&0\end{pmatrix}  \quad F^{3} = \frac{1}{2}\begin{pmatrix}1&&0&&0\cr0&&-1&&0\cr0&&0&&0\end{pmatrix} \quad,
\end{equation}
$$ F^{4} = \frac{1}{2}\begin{pmatrix}0&&0&&1\cr0&&0&&0\cr1&&0&&0\end{pmatrix} \quad F^{5} = \frac{1}{2}\begin{pmatrix}0&&0&&-i \cr 0&&0&&0\cr i&&0&&0\end{pmatrix}  \quad F^{6} = \frac{1}{2}\begin{pmatrix}0&&0&&0\cr0&&0&&1\cr0&&1&&0\end{pmatrix} \quad, $$
$$ F^{7} = \frac{1}{2}\begin{pmatrix}0&&0&&0\cr0&&0&&-i\cr0&&i&&0\end{pmatrix} \quad F^{8} = \frac{1}{2\sqrt{3}}\begin{pmatrix}1&&0&&0 \cr 0&&1&&0\cr 0&&0&&-2\end{pmatrix}   \quad. $$
By inspection, these $F^{i}$s are hermitian and traceless.

Notice that $\{F^{1},F^{2},F^{3}\}$ contain the  Pauli  $2 \times 2$ spin matrices of SU(2). Embedding standard matrices from SU(2) is a major simplification in the construction of SU(3) matrices for other irreps.

\section{General Considerations} \label{GenC}

The $(p,q)$ irrep, labeled with nonnegative integers $p$ and $q,$ of SU(3) can be represented with a basis of eight $d \times d$ square matrices $F^{i}$ of dimension $d$ = $(p+1)(q+1)(p+q+2)/2.$ \cite{GandG} The matrices in (\ref{Gell-Mann}) represent the $(p,q)$ = $(1,0)$ irrep with $d$ = 3 dimensions. 

Since the negative transpose of the eight basis matrices for the $(p,q)$ irrep is a representation for $(q,p),$ we need only consider $p\geq q$ and use the negative transpose for the $q>p$ irreps.

For a $(p,q)$ irrep, the matrix $D(\theta)$ in (\ref{expiaF}) is a transformation matrix acting on $d$-component `vectors' each of whose components is proportional to a `state'. The basis states are vectors each having just one nonzero component.\cite{GandG}  Two of the matrices, $T^{3}$ and $U^{3},$ are here diagonal and so the basis states are eigenvectors of $T^{3}$ and $U^{3}$ with eigenvalues making the diagonal components of the matrices $T^{3}$ and $U^{3}.$ The eigenvalues are also called `quantum numbers' and the basis states form a `multiplet'.

It is equivalent to work with the eight matrices $\{T^{+},T^{-},T^{3},V^{+},V^{-},U^{+},U^{-},U^{3}\},$ where
\begin{equation} \label{FtoTUV} T^{\pm} = F^{1} \pm i F^{2} \quad ; \quad T^{3} =  F^{3} \quad ; \quad V^{\pm} = F^{4} \pm i F^{5} \quad;
\end{equation}
$$ U^{\pm} = F^{6} \pm i F^{7} \quad ; \quad  U^{3} = - \frac{1}{2} F^{3} + \frac{\sqrt{3}}{2} F^{8}\quad . $$
These equations can be inverted to find the $F^{i}$s from the $T$s, $U$s and $V$s.

We require the $F^{i}$ to be (i) hermitian and (ii) traceless. Furthermore we insist that (iii) $\{F^{1},F^{3},F^{4},F^{6},F^{8}\}$ have real-valued components while (iv) $\{F^{2},F^{5},F^{7}\}$ are pure imaginary.

By (i), (iii), and (iv), it follows that (A) the $\{T^{+},T^{-},T^{3},V^{+},V^{-},U^{+},U^{-},U^{3}\}$ all have real-valued components. It also follows that 
 \begin{equation} \label{UmVm} {\mathrm{(B)}}\quad T^{-}_{ij} = T^{+}_{ji} \quad {\mathrm{and}} \quad {\mathrm{(C)}}\quad U^{-}_{ij} = U^{+}_{ji} \quad {\mathrm{and}} \quad {\mathrm{(D)}} \quad V^{-}_{ij} = V^{+}_{ji} \quad.
\end{equation}
Therefore, once we find $U^{+}$ and $V^{+}$ we get $U^{-}$ and $V^{-}$ by taking the transpose. We get $T^{-}$ another way.

The eight matrix basis has 28 = $8*7/2$ commutation relations, \cite{GandGandM}
 \begin{equation} \label{commRs1} [T^{3},T^{\pm}] = \pm  \, T^{\pm}  \quad ; \quad [T^{+},T^{-}] = 2 \, T^{3} \quad ; \quad 
\end{equation}
 \begin{equation} \label{commRs2} [T^{3},U^{\pm}] = \mp \, \frac{1}{2} U^{\pm}  \quad ; \quad [T^{3},V^{\pm}] = \pm \, \frac{1}{2}  V^{\pm} \quad ; \quad [U^{3},T^{\pm}] = \mp \, \frac{1}{2} T^{\pm}  \quad ;
\end{equation}
 \begin{equation} \label{commRs3}  \quad [U^{3},U^{\pm}] = \pm \, U^{\pm}   \quad ; \quad [U^{3},V^{\pm}] = \pm \, \frac{1}{2} V^{\pm}   \quad ; \quad [T^{3},U^{3}] = 0   \quad ;
\end{equation}
 \begin{equation} \label{commRs4} [T^{\pm},U^{\mp}] = [T^{\pm},V^{\pm}] = 0    \quad ; \quad [T^{\pm},U^{\pm}] = \pm \,  V^{\pm}   \quad ; \quad [T^{\pm},V^{\mp}] = \pm \,  U^{\mp} \quad ; 
\end{equation}
 \begin{equation} \label{commRs5} [U^{+},U^{-}] = 2 U^{3}  \quad ; \quad [V^{+},V^{-}] = 2 U^{3} + 2 T^{3} \quad ; \quad [V^{\pm},U^{\mp}] = \pm \,  T^{\pm}    \quad ; \quad [U^{\pm},V^{\pm}] =  0    \quad . 
\end{equation}
Notice that the $T$s (and $U$s and $V$s) satisfy the commutation relations of SU(2). Following convention, we build the three $T$s from standard spin matrices.  

We write the $T$s as the direct sum of irreps of SU(2) with each SU(2) irrep occupying a block along the diagonal with vanishing components outside those blocks. Standard spin matrices are available for these diagonal blocks, so the $T$s are determined once the list of spins is known. The list of T-spins can be found by conventional methods. Conventional methods also give the values of $U^{3}$ for the basis states.\cite{GandG} 

Therefore we consider $T^{\pm},$ $T^{3},$ and $U^{3}$ as given, even though some work is needed to find expressions for them. By (\ref{UmVm}), we have $U^{-}$ and $V^{-}$ as transposes of $U^{+}$ and $V^{+}.$ Thus the unknown matrices are $U^{+}$ and $V^{+}.$

\section{States and T-Spins} \label{Tspins}

We start with the states of the $(p,q)$ irrep to find the list of T-spins. 

Since any two linear combinations of $T^{3}$ and $U^{3}$ are diagonal and commute with one another, the states can be identified just as well by other quantum numbers. It is conventional to graph the states as in Fig.  1 where the coordinates are the $T^{3}$ eigenvalue and the value of $Y$ = $(4/3) U^{3}$ + $(2/3) T^{3}.$ Many states have the same $T^{3},Y$ coordinates and the points in Fig.  1 are labeled with the number of states at the point, the `multiplicity'.

The $T^{3},Y$ plot is useful. The number of spaces in the top row in Fig.  1 is $p$ = 5 and the number of spaces in the bottom row is $q$ = 3. And, since $[T^{\pm},Y]$ = 0, the action of the step-up and step down matrices $T^{\pm}$ is to move from state to state on the same row, $Y$ = constant. 

For example the row in Fig.  1 with $Y$ = $-4/3$ has 16 states. Of those, 7 states have multiplicity 1 or more, 5 states have multiplicity 2 or more, 3 states have multiplicity 3 or more and 1 state has multiplicity 4. One can infer the T-spins $s$ for the row by setting the number of states equal to $2s+1,$ so the set of T-spins for the states on that row are $s$ = $\{3,2,1,0\}.$ 

Looking at all the rows in Fig. 1, one finds the following collection of T-spins, doubled for convenience and arranged from small to large, for the $(5,3)$ irrep,
$$  \quad 2s = \{0,1,1,2,2,2,3,3,3,3,4,4,4,4,5,5,5,5,6,6,6,7,7,8\} \quad .$$
There are $(5+1)(3+1)$ = 24 double T-spins $2s$ listed for the $(5,3)$ irrep. 

The T-spins are grouped into `top cap', `bottom cap' and `middle'. For the $(5,3)$ irrep, we have
$$ {\mathrm{top}} \, {\mathrm{cap:}}\quad 2s = \{0,1,1,2,2,2\} \quad \, ; \,{\mathrm{middle:}} \quad 2s = \{3,3,3,3,4,4,4,4,5,5,5,5\} \quad,  $$
$$ {\mathrm{bottom}} \, {\mathrm{cap:}} \quad 2s = \{6,6,6,7,7,8\} \quad . $$
Notice the symmetry; there is one smallest ($2s$ = 0) and one largest ($2s$ = 8), two next largest at $2s$ = 1 and two next smallest at $2s$ = 7, etc.

One finds similar results for the general $(p,q)$ irrep. There are $(p+1)(q+1)$ T-spins $s$. For the $(p,q)$ irrep  and doubled to $2s,$ we have
 \begin{equation} \label{Ts1}  \quad 2s = \{0,1,1,...,q-1;q,...,p;p+1,...,p+q-1,p+q-1,p+q\} \quad [\, p \geq q \, ] \quad ,
\end{equation}
where the semicolons separate the top cap, middle and bottom cap. For the $(p,q)$ irrep, we have for the doubled T-spins $2s,$
\begin{equation} \label{Ts2} {\mathrm{top}} \, {\mathrm{cap:}}\quad 2s = \{0,\underbrace{1,1}_
{\mbox{2}},...,\underbrace{q-1,...,q-1}_
{\mbox{total of q terms}}\} \quad ,
\end{equation}
$$ {\mathrm{middle:}} \quad 2s = \{\underbrace{q,...,q}_
{\mbox{q+1}},\underbrace{q+1,...,q+1}_
{\mbox{q+1}},...,\underbrace{p-1,...,p-1}_
{\mbox{q+1}},\underbrace{p,...,p}_
{\mbox{q+1}}\} \quad , $$
$${\mathrm{bottom}} \, {\mathrm{cap:}} \quad 2s = \{\underbrace{p+1,...,p+1}_
{\mbox{q}},...,\underbrace{p+q-1,p+q-1}_
{\mbox{2}},p+q\}  \quad \quad [\, p \geq q \, ] \quad.  $$
For the $(q,p)$ irrep with $p > q,$ one finds the same T-spins, so the formulas (\ref{Ts2}) do not apply to any $(p,q)$ with $q > p.$ In the case of a $(p,q)$ irrep with $q > p,$ one would need to interchange $p$ and $q,$ $p \leftrightarrow q,$ in (\ref{Ts2}) to get the correct T-spin list.

Thus the number of T-spins $N_{T}$ in the list for the $(p,q)$ irrep is
$$ N_{T} = \sum^{q}_{i=1} i + (p-q+1)(q+1) + \sum^{q}_{k=1} k = (\frac{q}{2} + p-q + 1 + \frac{q}{2})(q+1) = (p+1)(q+1)  \quad .$$
Since the number of states for each T-spin $s$ is $2s+1,$ the number of states, i.e. the dimension $d$ of the $(p,q)$ irrep, is
$$ d = \sum^{q}_{i=1} i^{2} + (q+1)\sum^{p}_{j=q} (j+1) + \sum^{q}_{k=1} k(p+q-k+2) = \frac{1}{2}(p+1)(q+1)(p+q+2)  \quad .$$
This is the expected result \cite{GandG}, and confirms the list of T-spins.

\section{T-Matrices} \label{Tmats}

The basis matrices have $d$ dimensions. The components are separated into arrays called `blocks' based on T-spin.  The $ij$ block has $2s_{i}+1$ rows and $2t_{j}+1$ columns, where $s_{i}$ is the $i$th T-spin in the list (\ref{Ts1}) and $t_{j}$ is the $j$th T-spin in the list. There are $(p+1)(q+1)$ T-spins for the $(p,q)$ irrep, so its basis matrices have a $(p+1)(q+1)$ by $(p+1)(q+1)$ block structure.

Given the list of T-spins in Sec.  \ref{Tspins}, one can apply well-known standard spin matrices to obtain the T-matrices $T^{+},$ $T^{-},$ and $T^{3}.$ Arrange the states by the T-spin list (\ref{Ts1}) and (\ref{Ts2}) and within each T-spin put the state with the largest spin component first. This is the order in the Pauli matrices, larger spin component first. 

The T-matrices are block diagonal. The matrices $T^{+},$ $T^{-},$ and $T^{3}$ are given by \cite{WandEandR}
 \begin{equation} \label{Tmat1}  {T^{\pm}_{(i,j)}}_{\sigma \tau} = r^{\pm}(s_{j},\sigma) \delta_{i,j}\delta_{\sigma, \tau \pm 1} \quad ;  \quad {T^{3}_{(i,j)}}_{\sigma \tau} = \tau \delta_{i,j}\delta_{\sigma,\tau} \quad ,
\end{equation}
where $s_{j}$ is the $j$th T-spin in the list (\ref{Ts1}) and
\begin{equation} \label{rprm}r^{+}(s,\sigma) \equiv \sqrt{(s - \sigma)(s + \sigma +1) } \quad ; \quad r^{-}(s,\sigma) \equiv \sqrt{(s + \sigma)(s - \sigma +1) } \quad ,
\end{equation}
with $\sigma, \tau \in$ $\{s, s-1,...,-(s-1),-s\}$ and $\delta_{\sigma \tau}$ = 1 for $\sigma$ = $\tau$ and  $\delta_{\sigma \tau}$ = 0 otherwise. Any component of the T-matrices outside these diagonal blocks vanishes.

\section{$U^{3}$} \label{SecU3}

We assign addresses to components  the same as for the T-matrices: ${U^{3}_{(ij)}}_{\sigma \tau},$ with $\sigma \in$ $\{s_{i},s_{i}-1,...,-s_{i} \}$ and $\tau \in$ $\{t_{j},t_{j}-1,...,-t_{j} \},$ where $s_{i}$ and $t_{j}$ are the $i$th and $j$th spins in the T-spin list (\ref{Ts1}). Since $U^{3}$ is diagonal, we have 
 \begin{equation} \label{uc} {U^{3}_{(ij)}}_{\sigma \tau} = u^{3}(i,\sigma)\delta_{ij} \delta_{\sigma \tau} \quad .
\end{equation} 
Thus $u^{3}(i,s_{i})$ is the lead diagonal element of the $ii$ block of $U^{3}.$ 

The values of $u^{3}(i,\sigma)$ are the eigenvalues of $U^{3}$ for the basis states and can be determined from the values of $T^{3}$ and $Y$ obtained for a given $(p,q)$ by conventional means. For example, for the $(5,3)$ irrep in Fig. 1, for the state $\mid T^{3}, Y\rangle$ = $\mid 4,2/3\rangle,$ the right-most point on the plot, we have $u^{3}(24,4)$ = $3Y/4 - T^{3}/2$ = $-3/2$  .

For any $(p,q)$ irrep, the commutation relation $[U^{3},T^{+}]$ = $- 2T^{+}$ implies a recursion for the components in any one diagonal block. One finds
 \begin{equation} \label{uca}  {U^{3}_{(ii)}}_{\rho \sigma}{T^{+}_{(ii)}}_{ \sigma \tau} -{T^{+}_{(ii)}}_{\rho \sigma}{U^{3}_{(ii)}}_{ \sigma \tau} = -\frac{1}{2} {T^{+}_{(ii)}}_{\rho \sigma} \quad ,
\end{equation}
\vspace{.1cm}
$$ u^{3}(i,\sigma-1) = u^{3}(i,\sigma)+\frac{1}{2} \quad ; \quad u^{3}(i,\sigma) = u^{3}(i,s_{i})+\frac{s_{i}-\sigma}{2} \quad ,$$
where $s_{i}$ is the $i$th T-spin and there is no sum over $i.$
Thus we need only the lead component of each block $u^{3}(i,s_{i})$ to get all the $u^{3}(i,\sigma)$s for the block. 

For the $(p,q)$ = $(5,3)$ irrep, the lead components of the 24 diagonal blocks of $U^{3}$ are
$$ {\mathrm{top}} \, {\mathrm{cap:}}\quad 2u^{3}(i,s_{i}) = \{-2,-4,-1,-6,-3,0\} \quad , $$ 
$${\mathrm{middle:}} \quad 2u^{3}(i,s_{i}) = \{-8,-5,-2,1,-7,-4,-1,2,-6,-3,0,3\} \quad,  $$
$$ {\mathrm{bottom}} \, {\mathrm{cap:}} \quad 2u^{3}(i,s_{i}) = \{-5,-2,1,-4,-1,-3\} \quad , $$
where the lead components are doubled for convenience and the states are reordered if needed so that states with the same T-spin $s$ but lower lead value $u^{3}(i,s)$ are first.

In general, for the $(p,q)$ irrep, the lead components of the $(p+1)(q+1)$ diagonal blocks  of $U^{3}$ are found to be
\begin{equation} \label{uc1} {\mathrm{top}} \, {\mathrm{cap:}}\quad 2u^{3}(k,s_{k}) = -p+q-2i+3j \quad ,
\end{equation}
where $k$ = $1 + i(i+1)/2 +j,$ $i \in$ $\{0,1,...,q-1\}$ and $j \in$ $\{0,1,...,i-1\},$
$$ {\mathrm{middle:}} \quad 2u^{3}(k,s_{k}) = -p-q+i+3j \quad , $$
where $k$ = $q(q+1)/2+i(q+1)+j+1,$  $i \in$ $\{0,1,2,...,p-q\}$ and $j \in$ $\{0,1,2,...,q\},$
$${\mathrm{bottom}} \, {\mathrm{cap:}} \quad 2u^{3}(k,s_{k}) = -2q+i+3j+1   \quad \quad [\, p \geq q \, ] \quad,  $$
where $k$ = $(p+1)(q+1)-(q-i)(q-i+1)/2+j+1,$ $i \in$ $\{0,1,...,q-1\}$ and $j \in$ $\{0,1,...,q-1-i\}.$
The lead components are again doubled for display. 

For a $(q,p)$ irrep with $p > q,$ we have ${u^{3}}^{\prime}(k,s_{k})$ = $-{u^{3}}(k,s_{k}) -$ $s_{k}.$ Because of the minus sign, there is also a reordering of the states with the same $s_{k}.$ Instead of writing more expressions like (\ref{uc1}) for the case $(p,q)$ with $q > p,$ we use the negative transpose of the $(p,q)$ irrep to get matrices for the $(q,p)$ irrep with $p>q.$ 

At this point, for the $(p,q)$ irrep with $p \geq q,$ we have a one-to one labeling of the states. Besides $p$ and $q,$ we have put states with lower T-spin first and for a given T-spin $s_{i}$ states with larger 3-component $\sigma$ go first. This still leaves whole sets of $2s_{i}+1$ states each that can be interchanged. The sets of states with the same  T-spin $s_{i}$ are ordered by $U^{3},$ with a $2s_{i}+1$ state set with lower lead $u^{3}(i,s)$ is first. Thus a state can be identified by the five quantities $\{p,q,s^{2}_{i}, \sigma, u^{3}(i,\sigma)\}$ and the values of those quantities determines the  order of the states. 

Thus, for $p \geq q,$ we can determine four matrices $T^{\pm},$ $T^{3},$ and $U^{3}.$ It remains to find $U^{+}$ and $V^{+}.$

\section{$U^{+}$ and $V^{+}$} \label{UpVp}

To find the matrices $U^{+}$ and $V^{+}$ we first find that just one diagonal in each block can have nonzero components. Then we find which blocks can be nonzero. The nonzero blocks are off-diagonal, taking their places in the lessor diagonals above and below the main diagonal blocks. 

{\it{Inside a block.}} By the commutation relations $[T^{3},U^{+}]$ = $(-1/2)U^{+}$ and $[T^{3},V^{+}]$ = $(-1/2)V^{+}$ together with the expressions (\ref{Ts2}) for $T^{3},$ one finds that each $U^{+}$ or $V^{+}$ block can have but one nonzero diagonal,
\begin{equation} \label{UpVp1} {U^{+}_{(ij)}}_{\sigma \tau} = u^{+}_{(ij)}(\sigma)\delta_{\sigma,\tau-1/2} \quad {\mathrm{and}} \quad {V^{+}_{(ij)}}_{\sigma \tau} = v^{+}_{(ij)}(\sigma)\delta_{\sigma,\tau+1/2} \quad ,
\end{equation}
where $\sigma \in$ $\{s_{i},s_{i}-1,...,-s_{i} \},$  $\tau \in$ $\{t_{j},t_{j}-1,...,-t_{j} \},$ with $s_{i}$ and $t_{j}$ the $i$th and $j$th T-spin in the list (\ref{Ts1}). Since $\sigma$ and $\tau$ must differ by one half, if $2s_{i}$ is even in a nonzero block, then $2t_{j}$ is odd and vica versa.

Recursions follow from the commutation relations $[T^{-},U^{+}]$ = 0 and $[T^{+},V^{+}]$ = 0. One finds that
\begin{equation} \label{UpVp2} r^{-}(t_{j},\sigma+1/2)u^{+}_{(ij)}(\sigma-1) = r^{-}(s_{i},\sigma)u^{+}_{(ij)}(\sigma)  \quad ,
\end{equation}
for $u^{+}$ and
$$  r^{+}(s_{i},\sigma-1)v^{+}_{(ij)}(\sigma-1) = r^{+}(t_{j},\sigma-3/2)v^{+}_{(ij)}(\sigma)  \quad ,$$
for $v^{+}.$

Therefore, by (\ref{UpVp1}), the nonzero components of each block ${U^{+}_{(ij)}}$ and ${V^{+}_{(ij)}}$ form just one diagonal in the block and, by (\ref{UpVp2}) the components on that diagonal depend on the value of the lead component. There is just one unknown per block, one for ${U^{+}_{(ij)}}$ and one for ${V^{+}_{(ij)}}$. 

But, by the commutation relation $[T^{+},U^{+}]$ = $V^{+},$ the unknown quantity $v^{+}_{(ij)}$ of the ${V^{+}_{(ij)}}$ block is proportional to the unknown $u^{+}_{(ij)}$ for the ${U^{+}_{(ij)}}$ block, the factor depending on whether $s_{i}$ is greater than or less than $t_{j}.$  Thus we need to find one unknown quantity $u^{+}_{(ij)}$ per block.

{\it{Nonzero blocks of $U^{+}$ and $V^{+}$.}} 
We now use the diagonal matrix $U^{3}.$ From $[U^{3},U^{+}]$ = $ U^{+}$ and $[U^{3},V^{+}]$ = $ V^{+},$ one finds that the blocks ${U^{+}_{(ij)}}$ and ${V^{+}_{(ij)}}$ can be nonzero only for 
\begin{equation} \label{UpVp3} u^{3}(i,s_{i})-u^{3}(j,t_{j}) = \frac{3 \pm 1}{4} \quad {\mathrm{with}} \quad t_{j} = s_{i} \pm \frac{1}{2}   \quad \quad [\, p \geq q \, ]\quad ,
\end{equation}
where $u^{3}(i,s_{i})$ is the lead diagonal component of the $ii$-block of the $U^{3}$ matrix. The values of $2u^{3}(i,s_{i})$ are listed in (\ref{uc1}).

Since $s_{i}$ and $t_{j}$ must differ by $\pm 1/2,$ the recursions (\ref{UpVp1}) and (\ref{UpVp2}) give the expressions
\begin{equation} \label{UpREC} u^{+}_{(ij)}(\sigma) =  \sqrt{s_{i} - \sigma} \, \, u^{+}_{(ij)} \quad (t_{j} = s_{i} -1/2)  ; \quad u^{+}_{(ij)}(\sigma) =  \sqrt{ \frac{s_{i} + \sigma+1}{ 2 s_{i} +1} }\, \, u^{+}_{(ij)}  \quad (t_{j} = s_{i} +1/2) \, ,
\end{equation}
and 
\begin{equation} \label{VpREC}
v^{+}_{(ij)}(\sigma) = \sqrt{s_{i} + \sigma} \, \, u^{+}_{(ij)}  \quad (t_{j} = s_{i} -1/2) ; \quad  v^{+}_{(ij)}(\sigma) = -\sqrt{ \frac{s_{i} - \sigma+1}{ 2 s_{i} +1} }\, \, u^{+}_{(ij)} \quad (t_{j} = s_{i} +1/2) \, ,
\end{equation}
where $p \geq q$ and $u^{+}_{(ij)}$ is a constant, one for each block.

Roughly speaking, for $(p,q)$ = $(5,3),$ there is  a nonzero diagonal zig-zagging above and another one below the main diagonal which adds up to a few less than $2 \times 24$ possibly nonzero blocks in $U^{+}.$  Since the dimension $d$ of the $(5,3)$ irrep is 120, the problem of finding the $2 \times 120^{2}$ components of ${U^{+}}$ and ${V^{+}}$ is thus reduced to finding 40 or so values of $u^{+}_{(ij)}.$ We turn now to that.

\section{Finding the $U^{+}$ Block Unknowns} \label{ups}

One finds that the squares of the $u^{+}_{(ij)}$s are rational. By solving for the ${u^{+}_{(ij)}}^2$s, and then taking the square root to find the $u^{+}_{(ij)}$s, one introduces a number of arbitrary $\pm$ signs. Here we simply take the positive root which seems to work well.

The ${u^{+}_{(ij)}}^2$s form a $(p+1)(q+1)$ square matrix. The matrix reflects the top cap, middle, bottom cap structure of the list of T-spins for the states of the $(p,q)$ irrep, (\ref{Ts1}). One finds that, for any $(p,q)$ the nonzero blocks are located above and below the diagonal in each of the three regions. The ${u^{+}_{(ij)}}^2$ matrix has the general structure, for any $(p,q),$ 
\begin{equation} \label{tmb}
\begin{pmatrix}0&{\mathrm{utc}}&0&0&0&0\cr{\mathrm{ltc}}&0&0&0&0&0\cr0&0&0&{\mathrm{udm}}&0&0\cr0&0&{\mathrm{ldm}}&0&0&0\cr0&0&0&0&0&{\mathrm{ubc}}\cr0&0&0&0&{\mathrm{lbc}}&0\cr\end{pmatrix} \quad ,
\end{equation}
where we have {\it{utc}} - upper top cap, {\it{ltc}} - lower top cap, {\it{udm}} - upper diagonal middle, {\it{ldm}} - lower diagonal middle, {\it{ubc}} - upper bottom cap, {\it{lbc}} - lower bottom cap. The structure is evident in Fig. 2.

To find formulas for the ${u^{+}_{(ij)}}$s in the top cap and on the top diagonal of the middle region, a catalog of low-valued $(p,q)$ irreps was constructed by solving the commutation relations for each $(p,q)$ separately. For example, ${u^{+}_{(31)}}^{2}$ = $p(q+2)/2$ was found by this method.

Once the top cap and upper diagonal of the middle is obtained, the commutation relation $[V^{-},U^{+}]$ = $-T^{-}$ and the fact that $V^{-}$ is the transpose of $V^{+}$ produce the following equation for the ${u^{+}_{(ij)}}^{2}$s along the lower diagonal of the middle region,
\begin{equation} \label{BottDiag}
{u^{+}_{(j + q + 1, j)}}^{2} = -1+{u^{+}_{(j,\, j - q + 0 \, \,{\mathrm{ or }}\, -1)}}^{2}+\frac{1}{2s_{j}}{u^{+}_{(j-q+ 1 + 0\, \,{\mathrm{ or }}\, -1,j)}}^{2}-\frac{1}{2s_{j}+1}{u^{+}_{(j,j+q)}}^{2} \quad ,
\end{equation}
where we choose `0' in the central two terms when $j < n_{0}(q+1)$ and `$-1$' otherwise and $j \in$ $\{n_{0}(q) + 1 \,\,{\mathrm{ or }}\,\, 0,...,n_{0}(q)+(p-q+2)(q+1)-3 \},$  with $j$ starting with $n_{0}(q)+1$ for $q$ = 1 and starting with $n_{0}(q)$ for $q \geq 2.$ The quantity $s_{j}$ is the $j$th T-spin and 
\begin{equation} \label{n0}
n_{0}(q) \equiv \frac{q(q-1)}{2} +1 \quad .
\end{equation}
The terms in (\ref{BottDiag}) lie on a `+'-shape in the matrix because they are in the $j$th row or the $j$th column. Since expressions for the top cap and upper diagonal of the middle were previously found using the catalog, expressions for the lower diagonal of the middle portion follow from (\ref{BottDiag}).

With the top cap and middle done, the bottom cap was found by a combination of solving (\ref{BottDiag}) and by deducing expressions from the catalog of low $(p,q).$

\vspace{.2cm}

\noindent{\it{The ${u^{+}_{(i,j)}}^2$ formulas. }}
\vspace{.2cm}

We separate the formulas for $q$ = 0 because with $q$ = 0 there is no top cap, no upper diagonal for the middle and no bottom cap. There is just the lower diagonal for the middle (ldm) which is given by 
\begin{equation} \label{p0}  {\mathrm{ldm:}} \quad   {u^{+}_{(j+1,j)}}^{2} = p-j+1 \quad \quad [q=0] \quad ,\hspace{0cm}
\end{equation}
where $j \in$ $\{1,...,p\}.$
The result follows from (\ref{BottDiag}) because $n_{0}(0)$ = 1 and the last two terms on the right side vanish.

For $q \geq 1,$ one finds formulas for the top cap, 
\begin{equation} \label{upc2tc} {\mathrm{utc:}}  \quad {u^{+}_{(i,i+q_{1})}}^{2} =  \frac{n_{0}(q_{1}+1)-i}{q_{1}+1}(p+n_{0}(q_{1}+1)-i+1)(q-n_{0}(q_{1}+1)+i+1)  \quad \quad [q \geq 1] \quad ,
\end{equation}
where $q_{1} \in$ $\{1,2,...,q-1\},$ $i \in$ $\{n_{0}(q_{1}),...,n_{0}(q_{1}+1)\}.$ We have
$${\mathrm{ltc:}} \quad {u^{+}_{(j+q_{1}+1,j)}}^{2} =  \frac{j-n_{0}(q_{1})+1}{q_{1}(q_{1}+1)}(p+n_{0}(q_{1})-j)(q-n_{0}(q_{1})+j+2)   \quad , \hspace{0cm} $$
where $q_{1} \in$ $\{1,2,...,q-1\},$ $j \in$ $\{n_{0}(q_{1}),...,n_{0}(q_{1}+1)-1\}.$ In addition, 
$${\mathrm{ltc:}} \quad {u^{+}_{(3,1)}}^{2} = \frac{1}{2}p(q+2) \quad .\hspace{0cm}$$
The middle region ${u^{+}_{(ij)}}^{2}$s are found to be
\begin{equation} \label{upc2mid}  {\mathrm{udm:}} \quad {u^{+}_{(i,i+q)}}^{2} =  \frac{q- m(i,q)}{q+1+f(i,q)}(p+q+1-m(i,q))(1+m(i,q))    \quad , \hspace{0cm}
\end{equation}
where $i \in$ $\{n_{0}(q),...,N-q(q+1)/2\}$ and $$ m(i,q) \equiv i-n_{0}(q) \,\,\,{\mathrm{mod}}\,\,{ q+1} \quad ; \quad f(i,q) \equiv {\mathrm{floor}}(\frac{i-n_{0}(q)}{q+1})\quad . $$ We have
$$ {\mathrm{ldm:}} \quad {u^{+}_{(j+q+1,j)}}^{2} = -1+{u^{+}_{(j,j-q+0 \,\,{\mathrm{ or }}\,\, -1)}}^{2}+\frac{1}{2s}{u^{+}_{(j-q+1+0 \,\,{\mathrm{ or }}\,\, -1,j)}}^{2}-\frac{1}{2s+1}{u^{+}_{(j,j+q)}}^{2} \quad, \hspace{0cm} $$
where we choose `0' in the central two terms when $j < n_{0}(q+1)$ and `$-1$' otherwise and we have $j \in$ $\{n_{0}(q)+1 \,\,{\mathrm{ or }}\,\, 0,...,n_{0}(q)+(p-q+2)(q+1)-3\}$ with $j$ starting with $n_{0}(q)+1$ for $q$ = 1 and starting with $n_{0}(q)$ for $q \geq 2.$

For the bottom cap, we have
\begin{equation} \label{upc2bc}{\mathrm{ubc:}} \quad {u^{+}_{(i,i+q_{1})}}^{2} = (n_{0}(q_{1}+1)+i-N+q_{1}-1)(q-q_{1}-n_{0}(q_{1}+1)+N+2-i)\,\,\times 
\end{equation}
$$\hspace{0cm} \times \,\, \frac{p+q-q_{1}-n_{0}(q_{1}+1)+N+3-i}{p+q-q_{1}+2} \quad ,  $$
where $q_{1} \in$ $\{1,2,...,q-1\},$ $i \in$ $\{N-n_{0}(q_{1}+1)-q_{1}+1,...,N-n_{0}(q_{1})-q_{1}+1\}.$
$$ {\mathrm{lbc:}} \quad{u^{+}_{(j+q_{1}+1,j)}}^{2} =  (N-q_{1}-n_{0}(q_{1})+1-j)(p-N+q_{1}+n_{0}(q_{1})+j)\times $$ $$ \times \frac{p+q+j-N+q_{1}+n_{0}(q_{1})+1}{(p+q-q_{1}+1)(p+q-q_{1}+2)}   \quad,$$
where $q_{1} \in$ $\{1,2,...,q-1\},$ $j \in$ $\{N-n_{0}(q_{1}+1)+1-q_{1},...,N-n_{0}(q_{1})-q_{1}\}.$
These formulas complete the solution.

\section{Reconstruction} \label{build}

\noindent {\it{$(p,q)$ irrep with $p \geq q.$}}
\vspace{0.2cm} 

The formulas are written for a $(p,q)$ irrep with $p \geq q.$ The ${u^{+}_{(ij)}}^{2}$s in (\ref{p0}) - (\ref{upc2bc}) go into (\ref{UpREC}) and (\ref{VpREC}) to make $U^{+}$ and $V^{+}.$ Then $U^{-}$ and $V^{-}$ are the transpose matrices of $U^{+}$ and $V^{+}.$ The $T^{\pm},$ $T^{3},$ and $U^{3}$ are found in (\ref{Tmat1}), (\ref{uc}) and (\ref{uc1}). One now has all eight matrices $\{T^{\pm},T^{3},U^{\pm},U^{3},V^{\pm} \}.$ The eight $F^{i}$s can be found by (\ref{FtoTUV}).

\vspace{0.2cm}
\noindent {\it{$(p,q)$ irrep with $q > p.$}}
\vspace{0.2cm}

The formulas in the text give incorrect results when (mis)applied for $q > p.$ For example, the T-spins of the $(p,q)$ and $(q,p)$ irreps are the same. One can see this from a $T^{3},Y$ plot of the states like Fig. 1. For the $(p,q)$ irrep, the top row has $p$ spaces and there are $q$ spaces on the bottom row. The $(q,p)$ irrep has the same rows as but in different order with $q$ spaces on the top and $p$ spaces along the bottom row. Since the T-spin list depends on the rows and not their order, the T-spins are the same for $(p,q)$ and $(q,p).$ But (\ref{Ts1}) would give different T-spins for the two irreps. Thus we either write the correct formulas for $q > p$ or we try some other way. We go another way.

For $q > p$ we first calculate matrices as above for $(p^{\prime},q^{\prime})$ = $(q,p),$ since $p^{\prime}$ = $q >$ $p$ = $q^{\prime}.$ Then the negative transpose changes the $(p^{\prime},q^{\prime})$ matrices to an irrep for $(p,q).$ We have 
\begin{equation} \label{pqTOqp} \{T^{\pm}_{jk},T^{3}_{jk},V^{\pm}_{jk},U^{\pm}_{jk},U^{3}_{jk}\} = \{-{T^{\prime}}^{\pm}_{kj},-{T^{\prime}}^{3}_{kj},-{V^{\prime}}^{\pm}_{kj},-{U^{\prime}}^{\pm}_{kj},-{U^{\prime}}^{3}_{kj}\} \quad 
\end{equation}
and
\begin{equation} \label{Fherm} F^{i}_{jk} = -{F^{\prime}}^{i}_{kj} = -{{F^{\prime}}^{i}}^{\ast}_{jk} \quad ,
\end{equation}
where (\ref{FtoTUV}) gives the $F^{i}$ and since the ${{F^{\prime}}^{i}}$ are hermitian, the transpose of ${{F^{\prime}}^{i}}$ is equivalent to the complex conjugate of ${{F^{\prime}}^{i}}.$

When identifying the states, it is easier to work with the complex conjugate, $F^{i}_{jk}$ = $-{{F^{\prime}}^{i}}^{\ast}_{jk},$ because the order of the states is not changed. The states for $(p^{\prime},q^{\prime})$ are identified by the labels $\mid p^{\prime},q^{\prime}, s^{\prime}, \sigma^{\prime}, {u^{\prime}}^{3} \rangle.$ By definition, $(p^{\prime},q^{\prime})$ = $(q,p).$ Since $(p,q)$ and $(q,p)$ have the same list of T-spins, we have $s^{\prime}$ = $s.$ The minus signs in the real, diagonal matrices $F^{3}_{jk}$ = $-{{F^{\prime}}^{3}}^{\ast}_{jk}$ and $F^{8}_{jk}$ = $-{{F^{\prime}}^{8}}^{\ast}_{jk}$ give minus signs for $\sigma$ and $u^{3}.$ We have $\sigma^{\prime}$ = $-\sigma$ and $ {u^{\prime}}^{3}$ = $ -{u}^{3}.$ Relabeling the state $\mid p^{\prime},q^{\prime}, s^{\prime}, \sigma^{\prime}, {u^{\prime}}^{3} \rangle,$ we get  $\mid q,p,s, -\sigma, -{u}^{3} \rangle.$

We must also consider the order of the states to know what state $j$ is connected to which state $k$ by the matrix element $F^{i}_{jk}.$ By construction, the ordering in the matrices is the ordering of the primed system for $(p^{\prime},q^{\prime}).$ Since the T-spins are the same, the blocks have the same order of T-spins $s_{k}.$ But the minus sign $\sigma^{\prime}$ = $-\sigma$ reverses the order for the states with the $k$th T-spin $s_{k},$ so that these $2s_{k}+1$ states have 3-components $\sigma \in$ $\{-s_{k},...,s_{k}\},$ i.e. the lowest 3-component first which is just the reverse of the $p>q$ case. States with the same T-spins but higher values of $U^{3}$ occur first because of the minus sign in  $ {u^{\prime}}^{3}$ = $ -{u}^{3}.$ Further complicating matters, is the fact that the lead T-spin 3-component changes sign $s \rightarrow$ $-s.$ Then the lead component of the $kk$ block of $U^{3}$ has ${u^{3}}^{\prime}(k,s_{k})$ = $-{u^{3}}(k,s_{k}) -$ $s_{k}.$ Thus the labeling of the states and the order of states can be determined for the $(p,q)$ irrep with $q > p.$ 

\section{Mathematica Notebook} \label{mma}

This Mathematica \cite{Wolfram} notebook calculates the eight basis matrices $T^{\pm}$, $T^{3},$ $U^{\pm}$, $U^{3},$ and $V^{\pm}$ for the (p,q) irrep of the su(3) algebra. The notebook checks that the eight matrices obey the su(3) algebra and the quadratic Casimir invariance equation. The eight basis matrices $F^{i},$ $i \in$ $\{1,2,...,8\},$ are also calculated and the $F^{i}$ version of the quadratic Casimir invariance equation is checked. 

The Mathematica notebook was copied verbatim into Latex, turning $\sigma$ into Mathematica's `[Sigma]', {\textit{etc.}}. Exporting the pdf notebook to plain text may require minor changes, carriage returns and the like, to run. A link to the ready-to-run file is provided in Ref. \cite{notebook}. 

\pagebreak
\begin{verbatim}
(*Options*)
SetOptions[EvaluationNotebook[], ShowCellLabel -> False]
SetOptions[EvaluationNotebook[], PageWidth -> 540]

(*SU(3) Matrices,a notebook by R. Shurtleff May 2023 
Department of Applied Math and Sciences,Wentworth Institute of \
Technology,Boston MA 02115 
https://www.wolframcloud.com/obj/shurtleffr/Published/SU3MatricesMMA.\
nb 
https://www.dropbox.com/s/hpnuu46hawoh7i1/SU3MatricesMMA.nb?dl=0*)

(*This notebook calculates 8 basis matrices {Tp,Tm,T3,Up,Um,U3,Vp,Vm} \
(p is+and m is-) for the (p,q) irrep of the su(3) algebra.Notation \
follows S.Gasiorowicz,Elementary Particle Physics,(John \
Wiley\&Sons,Inc.,New York,1966),Equations (17.21-24)*)


(*This Mathematica notebook is a revision of the original notebook \
circa 2009. The current,May 18,2023,version runs with Mathematica \
13.0.1.0 on a Microsoft Windows (64-bit) platform.Wolfram \
Research,Inc.,Mathematica,Version 13.0,Champaign,IL (2022)*)

(*The notebook checks that the matrices found \
{Tp,Tm,T3,Up,Um,U3,Vp,Vm} obey the su(3) algebra and that they give \
the correct quadratic Casimer operator C1.The eight basis \
matricesF^iare also calculated.*)

(*The components of a basis matrix are organized into blocks along \
the diagonal.Tspins-list of SU(2) spins,from small to large.Tspin \
matrix irreps determines the block structure of the matrices.The \
T-spin list has structure:{top cap,middle,bottom cap}.numTspins-the \
number of spins in the list Tspins.block-The ij block is a matrix \
with 2Subscript[s, i]+1 rows and 2Subscript[t, j]+1 columns where s_i \
is the ith T-spin and t_j is the jth T-spin*)

(*U3firsts-U3 is a diagonal matrix.U3firsts is the list of the first \
value of U3 in each block.Ys-the list of the first Ys in each \
block.The states have U3=3Y/4-T3/2. (Y plays no part in the \
construction of the basis matrices.)*)

(*rp and rm-standard SU(2) functions;p stands for+(plus) and m stands \
for-(minus).{Tplus,Tminus,Tthree,Uplus,Uminus,Uthree,Vplus,Vminus} \
are block by block formulas.{Tp,Tm,T3,Up,Um,U3,Vp,Vm} are the \
matrices sought.*)

(*upc[i,j]-unknowns.One upc[i,j] for each block.The unknowns upc[i,j] \
are most directly related to Up.'uppertopcap' and'lowertopcap' are \
lists of formulas for the top cap portion of the upc2[i,j] \
matrix,where upc2[i,j]=(upc[i,j]^2)'upperdiagonal' and'lowerdiagonal' \
are the lists of upc2[i,j] formulas for the middle portion of the \
upc2[i,j] matrix
'upperbottomcap' and'lowerbottomcap' are the lists of upc2[i,j] \
formulas for the bottom cap portion of the upc2[i,j] matrix
There are miscellaneous values of upc2[i,j] that are introduced to \
cover special cases such as q=0,1.*)

(*Begin: 
Choose (p,q):*)

p = 3; q = 2;
If[p >= q, pqOK = True, pqOK = False];
If[pqOK, "sweet", p0 = p];
If[pqOK, "sweet", q0 = q];
If[pqOK, "sweet", p = q0];
If[pqOK, "sweet", q = p0];
If[pqOK, Print[
  "The formulas are set up for p\[GreaterEqual]q - no problem"], 
 Print["The formulas are set up for p\[GreaterEqual]q, so the \
notebook calculates with {p,q} = ", {p, q}, 
  " and then takes the negative transpose to get the matrices for \
{p0,q0} = ", {p0, q0}]]

numTspins = (p + 1) (q + 1);
dimREP = 1/2 (p + 1) (q + 1) (p + q + 2);

(*Useful functions*)
n0[q_] := n0[q] = (q (q - 1) + 2)/2;
mod[n_, q_] := mod[n, q] = Mod[n - n0[q], q + 1];
floor[n_, q_] := floor[n, q] = Floor[(n - n0[q])/(q + 1)];

(*Standard Spin functions: *)
rp[s_, \[Sigma]_] := 
  rp[s, \[Sigma]] = Sqrt[(s - \[Sigma]) (s + \[Sigma] + 1)];
rm[s_, \[Sigma]_] := 
  rm[s, \[Sigma]] = Sqrt[(s + \[Sigma]) (s - \[Sigma] + 1)];

(*Lists of values unique to each block.*)Tspins =(*The list of SU(2) \
spins the diagonal blocks of the T matrices.*)(1/
    2) Flatten[{Table[s2 - 1, {s2, q}, {m, s2}], 
    Table[s2, {s2, q, p}, {m, q + 1}], 
    Table[s2, {s2, p + 1, p + q}, {m, 
      q + 1 - s2 + 
       p}]}]; Yfirsts =              (* Yfirsts is not used in the \
construction.*)
 Flatten[{Table[-2 (p - q) - 3 ( i - 1) + 6 (j - 1), {i, q}, {j, i}], 
   Table[Table[-2 p - q + 3 i + 6 j, {j, 0, q}], {i, 0, p - q}],
   Table[3 + p - 4 q + 3 (i - 1) + 6 (j - 1), {i, q}, {j, 
     q - i + 1}]}]; U3firsts =   (*The list of first diagonal \
components of the blocks of the U3 matrix.*)(1/
    2) Flatten[{Table[-(p - q) - 2 (i - 1) + 3 (j - 1), {i, q}, {j, 
      i}], Table[Table[(-p - q + i + 3 j), {j, 0, q}], {i, 0, p - q}],
    Table[(1 - 2 q + (i - 1) + 3 (j - 1)), {i, q}, {j, q - i + 1}]}];

(*Standard SU(2) spin matrices. Step-up, step-down, diagonal.*)
Tplus[s_] := 
  Tplus[s] = 
   Table[If[\[Sigma] == \[Sigma]1 + 1, rp[s, \[Sigma]1], 0, 
     ah], {\[Sigma], s, -s, -1}, {\[Sigma]1, s, -s, -1}];
Tminus[s_] := 
  Tminus[s] = 
   Table[If[\[Sigma] == \[Sigma]1 - 1, rm[s, \[Sigma]1], 0, 
     ah], {\[Sigma], s, -s, -1}, {\[Sigma]1, s, -s, -1}];
Tthree[s_] := 
  Tthree[s] = 
   Table[If[\[Sigma] == \[Sigma]1, \[Sigma], 0, ah], {\[Sigma], 
     s, -s, -1}, {\[Sigma]1, s, -s, -1}];
(*Print["With s = 1, for example, we have {Tplus,Tminus,Tthree} = \
",{Tplus[1]//MatrixForm,Tminus[1]//MatrixForm,Tthree[1]//MatrixForm},
" ."]
*)

(*Make the Tp,Tm,T3 matrices by placing standard spin matrices as \
diagonal blocks:*)
Tp = Table[0, {i, dimREP}, {j, dimREP}]; For[s = 1, s <= numTspins, s++,
  Table[Tp[[i + Sum[2 Tspins[[si]] + 1, {si, s - 1}], 
    j + Sum[2 Tspins[[si]] + 1, {si, s - 1}]]] = 
   Tplus[Tspins[[s]]][[i, j]], {i, 2 Tspins[[s]] + 1}, {j, 
   2 Tspins[[s]] + 1}]]; Tm = Table[0, {i, dimREP}, {j, dimREP}];
For[s = 1, s <= numTspins, s++, 
 Table[Tm[[i + Sum[2 Tspins[[si]] + 1, {si, s - 1}], 
    j + Sum[2 Tspins[[si]] + 1, {si, s - 1}]]] = 
   Tminus[Tspins[[s]]][[i, j]], {i, 2 Tspins[[s]] + 1}, {j, 
   2 Tspins[[s]] + 1}]]; T3 = Table[0, {i, dimREP}, {j, dimREP}]; For[
 s = 1, s <= numTspins, s++, 
 Table[T3[[i + Sum[2 Tspins[[si]] + 1, {si, s - 1}], 
    j + Sum[2 Tspins[[si]] + 1, {si, s - 1}]]] = 
   Tthree[Tspins[[s]]][[i, j]], {i, 2 Tspins[[s]] + 1}, {j, 
   2 Tspins[[s]] + 1}]];

(* U3: *)
U3 = Table[0, {i, dimREP}, {j, dimREP}];(*Initialize*)
For[s = 1, s <= numTspins, s++, 
  Table[U3[[i + Sum[2 Tspins[[si]] + 1, {si, s - 1}], 
     i + Sum[2 Tspins[[si]] + 1, {si, s - 1}]]] = 
    U3firsts[[s]] + (1/2) (i - 1), {i, 2 Tspins[[s]] + 1}]];

(*Block unknowns upc and upc2 = upc^2 *)
(*Initialize*)
For[i = 0, i <= numTspins, i++, For[j = 0, j <= numTspins, j++,
   upc2[i, j] = 0; upc[i, j] = 0
   ]];
(*Miscellaneous*)
If[q == 1, upc2[3, 1] = p (q + 2)/2; upc[3, 1] = upc2[3, 1]^(1/2)];
(*uppertopcap*)
For[q1 = 1, q1 <= q - 1, q1++, For[i = n0[q1], i <= n0[q1 + 1], i++,
   upc2[i, 
     i + q1] = ((n0[q1 + 1] - i)/(q1 + 1)) (p + (n0[q1 + 1] - i) + 
       1) (q - (n0[q1 + 1] - i) + 1);   
   upc[i, i + q1] = upc2[i, i + q1]^(1/2)
   ]];
(*lowertopcap*)
For[q1 = 1, q1 <= q - 1, q1++, For[j = n0[q1], j <= n0[q1 + 1] - 1, j++,
   upc2[j + q1 + 1, 
     j] = ((j - n0[q1]) + 
       1) (p - (j - n0[q1])) (q + (j - n0[q1]) + 2)/(q1 (q1 + 1));
   upc[j + q1 + 1, j] = upc2[j + q1 + 1, j]^(1/2)
   ]];
(*upperdiagonal*)
For[i = n0[q], i <= numTspins - (q (q + 1))/2, i++,
  upc2[i, 
    i + q] = (q - mod[i, q]) (p + q + 
      1 + (-1) mod[i, q]) (1 + mod[i, q])/(q + 1 + floor[i, q]);
  upc[i, i + q] = upc2[i, i + q]^(1/2)
  ];
(*upperbottomcap*)
For[q1 = 1, q1 <= q - 1, q1++, 
  For[i = numTspins - n0[q1 + 1] - q1 + 1, 
   i <= numTspins - n0[q1] - q1 + 1, i++,
   upc2[i, 
     i + q1] = ((n0[q1 + 1] + i - numTspins + q1 - 1) (q - q1 - 
         n0[q1 + 1] + numTspins + 2 - i) (p + q - q1 - n0[q1 + 1] + 
         numTspins + 3 - i))/(p + q - q1 + 2); 
   upc[i, i + q1] = upc2[i, i + q1]^(1/2)
   ]];
(*lowerbottomcap*)
For[q1 = 1, q1 <= q - 1, q1++, 
  For[j = numTspins - n0[q1 + 1] + 1 - q1, 
   j <= numTspins - n0[q1] - q1, j++,
   upc2[j + q1 + 1, 
     j] = ((numTspins - q1 - n0[q1] + 1 - j) (p - numTspins + q1 + 
         n0[q1] + j) (p + q + j - numTspins + q1 + n0[q1] + 1))/((p + 
         q - q1 + 1) (p + q - q1 + 2)); 
   upc[j + q1 + 1, j] = upc2[j + q1 + 1, j]^(1/2)
   ]];
(*lowerdiagonal*)
If[q == 1, ifq = 1, ifq = 0];
If[q >= 1, 
  For[j = n0[q] + ifq, j <= n0[q] + (p - q + 2) (q + 1) - 3, j++, 
   If[j < n0[q + 1], ifj = 0, ifj = -1]; 
   upc2[j + q + 1, j] = -1 + upc2[j, j - q + ifj] + 
     upc2[j - q + 1 + ifj, 
       j]/(2 Tspins[[j]]) + (-1) upc2[j, j + q]/(2 Tspins[[j]] + 1); 
   upc[j + q + 1, j] = upc2[j + q + 1, j]^(1/2)
   ], For[i = 1, i <= p, i++, upc2[i + 1, i] = p - i + 1; 
   upc[i + 1, i] = upc2[i + 1, i]^(1/2)
   ]];

Unprotect[Up];

(* Up:*)
(*Tspins[[s]],Tspins[[t]] are SU(2) spins, one integral, one half \
integral.*)
Uplus[s_, t_] := 
  Table[If[\[Sigma] == \[Rho] - 1/2, 
    If[Tspins[[s]] > Tspins[[t]], 
     upc[s, t] Sqrt[Tspins[[s]] - \[Sigma]], 
     If[Tspins[[s]] < Tspins[[t]], 
      upc[s, t] Sqrt[ ((1 + Tspins[[s]] + \[Sigma])/(
       1 + 2 Tspins[[s]]))], ahu], huhu], 0, ah], {\[Sigma], 
    Tspins[[s]], -Tspins[[s]], -1}, {\[Rho], 
    Tspins[[t]], -Tspins[[t]], -1}];
Up = Table[0, {i, dimREP}, {j, dimREP}];
For[s = 1, s <= numTspins, s++, 
  For[t = 1, t <= numTspins, t++, 
   If[(Tspins[[t]] == 
       Tspins[[s]] + 1/2) && (U3firsts[[t]] - U3firsts[[s]] == -1), 
    Table[Up[[i + Sum[2 Tspins[[si]] + 1, {si, s - 1}], 
       j + Sum[2 Tspins[[ti]] + 1, {ti, t - 1}]]] = 
      Uplus[s, t][[i, j]], {i, 2 Tspins[[s]] + 1}, {j, 
      2 Tspins[[t]] + 1}]]]];
For[s = 1, s <= numTspins, s++, 
  For[t = 1, t <= numTspins, t++, 
   If[(Tspins[[t]] == 
       Tspins[[s]] - 1/2) && (U3firsts[[s]] - U3firsts[[t]] == 1/2), 
    Table[Up[[i + Sum[2 Tspins[[si]] + 1, {si, s - 1}], 
       j + Sum[2 Tspins[[ti]] + 1, {ti, t - 1}]]] = 
      Uplus[s, t][[i, j]], {i, 2 Tspins[[s]] + 1}, {j, 
      2 Tspins[[t]] + 1}]]]];

(* Vp:*)
(*Tspins[[s]],Tspins[[t]] are SU(2) spins, one integral, one half \
integral.*)
Vplus[s_, t_] := 
  Table[If[\[Sigma] == \[Rho] + 1/2, 
    If[Tspins[[s]] > Tspins[[t]], 
     upc[s, t] Sqrt[Tspins[[s]] + \[Sigma]], 
     If[Tspins[[s]] < Tspins[[t]], -upc[s, t] Sqrt[ ((
       1 + Tspins[[s]] - \[Sigma])/(1 + 2 Tspins[[s]]))], ahu], huhu],
     0, ah], {\[Sigma], Tspins[[s]], -Tspins[[s]], -1}, {\[Rho], 
    Tspins[[t]], -Tspins[[t]], -1}];
Vp = Table[0, {i, dimREP}, {j, dimREP}];
For[s = 1, s <= numTspins, s++, 
  For[t = 1, t <= numTspins, t++, 
   If[(Tspins[[t]] == 
       Tspins[[s]] + 1/2) && (U3firsts[[t]] - U3firsts[[s]] == -1), 
    Table[Vp[[i + Sum[2 Tspins[[si]] + 1, {si, s - 1}], 
       j + Sum[2 Tspins[[ti]] + 1, {ti, t - 1}]]] = 
      Vplus[s, t][[i, j]], {i, 2 Tspins[[s]] + 1}, {j, 
      2 Tspins[[t]] + 1}]]]];
For[s = 1, s <= numTspins, s++, 
  For[t = 1, t <= numTspins, t++, 
   If[(Tspins[[t]] == 
       Tspins[[s]] - 1/2) && (U3firsts[[s]] - U3firsts[[t]] == 1/2), 
    Table[Vp[[i + Sum[2 Tspins[[si]] + 1, {si, s - 1}], 
       j + Sum[2 Tspins[[ti]] + 1, {ti, t - 1}]]] = 
      Vplus[s, t][[i, j]], {i, 2 Tspins[[s]] + 1}, {j, 
      2 Tspins[[t]] + 1}]]]];

(*Um and Vm *)
Um = Table[0, {i, dimREP}, {j, dimREP}]; Table[
 Um[[i, j]] = Up[[j, i]], {i, dimREP}, {j, dimREP}]; Vm = 
 Table[0, {i, dimREP}, {j, dimREP}]; Table[
 Vm[[i, j]] = Vp[[j, i]], {i, dimREP}, {j, dimREP}];

(*For the case q>p :*)
If[pqOK, "sweet", Tp = -Transpose[Tp]];
If[pqOK, "sweet", Tm = -Transpose[Tm]];
If[pqOK, "sweet", T3 = -Transpose[T3]];
If[pqOK, "sweet", Up = -Transpose[Up]];
If[pqOK, "sweet", Um = -Transpose[Um]];
If[pqOK, "sweet", U3 = -Transpose[U3]];
If[pqOK, "sweet", Vp = -Transpose[Vp]];
If[pqOK, "sweet", Vm = -Transpose[Vm]];
If[pqOK, "sweet", p = p0];
If[pqOK, "sweet", q = q0];

(*Check:*)
Print["Check 29 equations for {p,q} = ", {p, q}, 
  " . All 29 eqns are satisfied: ", {0} == 
   Union[Flatten[
     Simplify[{T3 . Tp - Tp . T3 - Tp, T3 . Tm - Tm . T3 - (-Tm), 
       T3 . Up - Up . T3 - (-(1/2) Up), 
       T3 . Um - Um . T3 - (+(1/2) Um), T3 . U3 - U3 . T3, 
       T3 . Vp - Vp . T3 - 1/2 Vp, T3 . Vm - Vm . T3 - (-(1/2) Vm), 
       Tp . Tm - Tm . Tp - 2 T3, Tp . Up - Up . Tp - Vp, 
       Tp . Um - Um . Tp, Tp . U3 - U3 . Tp - (+(1/2) Tp), 
       Tp . Vp - Vp . Tp, Tp . Vm - Vm . Tp - (-Um), 
       Tm . Up - Up . Tm, Tm . Um - Um . Tm - (-Vm), 
       Tm . U3 - U3 . Tm - (-(1/2) Tm), Tm . Vp - Vp . Tm - Up, 
       Tm . Vm - Vm . Tm, U3 . Up - Up . U3 - (+ Up), 
       U3 . Um - Um . U3 - (- Um), U3 . Vp - Vp . U3 - (+(1/2) Vp), 
       U3 . Vm - Vm . U3 - (-(1/2) Vm), Up . Um - Um . Up - (2 U3), 
       Up . Vp - Vp . Up, Up . Vm - Vm . Up - (+ Tm), 
       Um . Vp - Vp . Um - (- Tp), Um . Vm - Vm . Um, 
       Vp . Vm - Vm . Vp - (2 U3 + 2 T3), (1/2) (Tp . Tm + Tm . Tp) + 
        T3 . T3 + (1/2) (Vp . Vm + Vm . Vp) + (1/2) (Up . Um + 
           Um . Up) + (1/
           3) (2 U3 + T3) . (2 U3 + 
            T3) - (((p^2 + p*q + q^2 + 3 p + 3 q)/3) IdentityMatrix[
           dimREP])}]]]];

(*The F^i basis matrices *)
Fi = {(1/2) (Tp + Tm), (-I/2) (Tp - Tm), 
   T3, (1/2) (Vp + Vm), (-I/2) (Vp - Vm), (1/2) (Up + Um), (-I/
      2) (Up + (-Um)), (2 U3 + T3)/Sqrt[3]};

Print["The F^i are traceless: ", {0} == 
   Union[Table[Sum[Fi[[i]][[j, j]], {j, dimREP}], {i, 8}]] ];
Print["and hermitian: ", {0} == 
   Union[Flatten[
     Table[Union[
       Flatten[Table[
         Fi[[i]][[j, k]] - Conjugate[Fi[[i]][[k, j]]], {j, 
          dimREP}, {k, dimREP}]]], {i, 8}]]]];

Print["The quadratic invariance can be written with the F^i matrices \
as F^2 = Sigma_i F^i.F^i = (p^2+p*q+q^2+3p+3q)/3 : ", {0} ==
   Union[Flatten[
     Simplify[
       Sum[Fi[[i]] . Fi[[i]], {i, 
         8}]] - (((p^2 + p*q + q^2 + 3 p + 3 q)/3) IdentityMatrix[
         dimREP])]]];

Print["The notebook found basis matrices {Tp,Tm,T3,Up,Um,U3,Vp,Vm} \
for the ", {p, q}, "-irrep."];
Print["The notebook also found the F^i, i = 1,...,8, basis matrices \
for the ", {p, q}, "-irrep."];

(*SetDirectory[NotebookDirectory[]];*)
(*Put[{Tp,Tm,T3,Up,Um,U3,Vp,Vm},"MatrixUploadFromMMAnotebook.dat"];*)
\end{verbatim}

\section{Fortran Program} \label{fortran}

The program calculates the eight basis matrices $T^{\pm}$, $T^{3},$ $U^{\pm}$, $U^{3},$ and $V^{\pm}.$ Writing the code was facilitated by running Code::Blocks 20.03 and the program was compiled with a GNU fortran compiler, Ref. \cite{fortran90}. 

\begin{verbatim}
!Calculating basis matrices for irreps of the SU(3) Lie algebra with Fortran 90
!   by Richard Shurtleff,
!   Wentworth Institute of Technology, Boston, MA, USA - retired
!	   email: shurtleffr(at)wit.edu, momentummatrix(at)yahoo.com
!
!   When working with a group, it is sometimes convenient to have numerical
!   matrix representations. Explicit formulas for basis matrices of SU(3) have
!   been available for decades.[1] However the earlier versions anyway are not
!   written in the language of currently popular elementary particle physics
!   texts.[2] Given the volume of literature since SU(3) became important in
!   particle physics, it is likely that formulas are available in the popular
!   language and have been coded into computer programs. However, the author
!   is unaware of any. This computer program builds the eight basis matrices
!   Tp,Tm,T3,Up,Um,U3,Vp,Vm of the finite dimensional (p0,q0)-irreducible
!   representation of SU(3), consistent with the notation and quantities found
!   in popular textbooks.[2]
!
!   License: CC BY-SA. Public use and modification of this code are allowed
!   provided that the preprint[3] or any subsequent published version is cited.
!
!   References
!   [1] See, for example, and references therein: L. C. Biedenharn, J. Math.
!   Phys. {\bf{4}}, 436 (1963);
!   G.E. Baird and L. C. Biedenharn, J. Math. Phys. {\bf{4}}, 1449 (1963);
!   {\bf{5}}, 1723 (1964);  {\bf{5}}, 1730 (1964).
!   [2] S. Gasiorowicz, Elementary Particle Physics, (John Wiley & Sons, Inc.,
!   New York, 1966), especially Equations (17.21 - 17.24). Also see W. Greiner
!   and B. Müller, Quantum Mechanics, Symmetries (Springer-Verlag, Berlin,
!   1994). Many other references use similar notations.
!   [3] R. Shurtleff, "Formulas for SU(3) Matrices," arXiv:0908.3864v2
!   [math-ph] 13 Mar 2023
!   [4] S. Weinberg, Quantum Theory of Fields Vol. I, Sec. 5.6 (1995).
!   [5] This program can be downloaded following the link:
!     https://www.dropbox.com/s/6tn7gnx8zph56ea/20230420SU3Formulas1e.f90?dl=0
!
!  The author welcomes all information and/or comments in regard to this code
!
!
!   For a link to the FORTRAN program file, see Ref. [5]
!
!------- CONTENTS -----------------------
!---1. Interface Blocks -----------------------
!---2. Type declarations -----------------------
!---3. Begin -----------------------
!---4. Preliminaries -----------------------
!---5. Make the Tp, Tm, T3, U3 matrices -----------------------
!---6. Make the Up, Vp, Um, Vm matrices -----------------------
!---7. Check that all matrix components are numbers, no 'NaN' ----------------
!---8. Check 28 commutator relations and an invariance equation --------------
!---9. Save the results to a file, end program -----------------------
!--10. External functions, end-of-file -----------------------
!
!INPUT p0 and q0 in Sec. 3 Begin. Also, one can reset the tolerance for error 
!   in the commutation relations and the quadratic Casimir equation. With single
!   precision arithmetic, I set the tolerance at MaxErrLimit = 1.E-4, which
!   is a value that I might decide to change and not tell you. Check the code.
!
!  CONSOLE OUTPUT
! Some data is displayed on the standard output. This can include the integers
!   p0 and q0 that identify the irrep, the dimension of the irrep's matrices,
!   the error tolerance, the maximum error found in 28 commutation relations
!plus the quadratic invariance, and other quantities deemed worth for display.
!  OUTPUT DATA File
! When the program succeeds in finding the matrix rep, a data file is created.
!   First line of OUTPUT: the integers p0, q0, the max error found in the 29
!   su(3) algebra equations, the 28 commutation relations plus the invariance
!   relation. Format: The format statement is FORMAT(2I3,1E16.7).
!
!   Second and following lines of OUTPUT: the components of the eight matrix
!   generators. The order of the components is:
!   Tp(1,1), Tp(1,2),Tp(1,3),...,Tp(1,dimREP),TP(2,1),TP(2,2),Tp(2,3),...,
!   Tp(dimREP,dimREP), then Tm(1,1),Tm(1,2),...,Tm(dimREP,dimREP),then
!  components in the same order for the rest of the matrices T3,Up,Um,U3,Vp,Vm
!  Format: The matrix components' format statement is FORMAT(5E16.7,/), so 5
!   real numbers per line of output.
!
!  SAMPLE DATA FILE
!  The data file for p0 = 2 an q0 = 1 reads, in part,
!
! 2  1   0.4768372E-06
! 0.0000000E+00   0.0000000E+00   0.0000000E+00   0.0000000E+00  0.0000000E+00
!  ...
!-0.1000000E+01   0.0000000E+00   0.0000000E+00   0.0000000E+00  0.0000000E+00
! end-of-datafile
!
!   Since the (p0,q0) = (2,1)-irrep has dimension dimREP = 15, the data shown
!   tells us that p0 = 2, q0 = 1, MaxErr = 0.4768372E-06, and Tp(1,1)=Tp(1,2)=
!   Tp(1,3)=Tp(1,4)=Tp(1,5) = 0. The last row of 5 numbers reads
!   Vm(15,11)=-0.01,Vm(15,12)=Vm(15,13)=Vm(15,14)=Vm(15,15)=0.
!   Note:For large matrices, one can save space by outputting only the nonzero
!   components.
!
!
PROGRAM SU3Formulas
!
    IMPLICIT NONE
!------------------------1. Interface Blocks -----------------------
INTERFACE               ! n0 = 1 + q(q-1)/2
    FUNCTION n0(q) RESULT  (w) !
        INTEGER :: w    ! n0 - often used in dummy indice limits.
        INTEGER, INTENT(IN) :: q  ! q characterizes the (p,q)-irrep
    END FUNCTION
END INTERFACE
INTERFACE               ! modA(n,q) = MOD(n-n0(q),q+1)
    FUNCTION modA(n,q) RESULT  (w)
        INTEGER :: w       ! modA(n,q) - often used in dummy indice limits.
        INTEGER, INTENT(IN) :: n,q  ! irrep is (p,q), where p,q are integers
    END FUNCTION
END INTERFACE
INTERFACE                   ! floorA(n,q) = FLOOR( (n-n0(q))/(q+1) )
    FUNCTION floorA(n,q) RESULT  (w) !
        INTEGER :: w        ! floor(n,q) - often used in dummy indice limits.
        INTEGER, INTENT(IN) :: n,q  ! irrep is (p,q), n >= n0(q)
    END FUNCTION
END INTERFACE
INTERFACE           ! standard spin formula,
    FUNCTION rp(s,sigma) RESULT  (w) ! S. Weinberg, Ref. [4], Eq. (5.6.17)
        REAL :: w                       !
        REAL, INTENT(IN) :: s,sigma  ! spin, spin component
    END FUNCTION
END INTERFACE
INTERFACE          ! standard spin formula,
    FUNCTION rm(s,sigma) RESULT  (w) ! S. Weinberg, Ref. [4], Eq. (5.6.17)
        REAL :: w
        REAL, INTENT(IN) :: s,sigma  ! spin, spin component
    END FUNCTION
END INTERFACE
INTERFACE              !Tplus(s) standard spin step up matrix, S. Weinberg [4]
    FUNCTION Tplus(s) RESULT  (w)
        REAL, INTENT(IN) :: s   ! the spin
      REAL :: sigma,sigma1,rp     ! spin components, spin function rp(s,sigma)
        REAL :: w(NINT(2*s+1),NINT(2*s+1))
        INTEGER :: i,j,sx2 ! two times the spin is an integer
    END FUNCTION
END INTERFACE
INTERFACE          !Tminus(s) standard spin step down matrix, S. Weinberg, [4]
    FUNCTION Tminus(s) RESULT  (w)
        REAL, INTENT(IN) :: s   ! the spin
        REAL :: sigma,sigma1,rm   ! spin components, spin function rm(s,sigma)
        REAL :: w(NINT(2*s+1),NINT(2*s+1))
        INTEGER ::  i,j,sx2  ! two times the spin is an integer
    END FUNCTION
END INTERFACE
INTERFACE              !Tthree(s), standard spin matrix,
    FUNCTION Tthree(s) RESULT  (w)      !  S. Weinberg, Ref. [4], Eq. (5.6.16)
        REAL, INTENT(IN) :: s   ! the spin
        REAL :: sigma,sigma1       ! spin components
        REAL :: w(NINT(2*s+1),NINT(2*s+1))
        INTEGER :: i,j,sx2     ! two times the spin is an integer
    END FUNCTION
END INTERFACE
INTERFACE   !The nth block starts at the end of the first n-1 Tspin blocks
    FUNCTION sumTmatrixDims(n,numTspins,Tspins) RESULT  (w)
        INTEGER, INTENT(IN) :: n,numTspins   ! nth spin, number of Tspins
        REAL, INTENT(IN) :: Tspins(numTspins)   ! the list of Tspins
        INTEGER :: i,k,w   ! dummy indices, the result is an integer
        REAL :: TspinN    ! dummy spin
    END FUNCTION
END INTERFACE
INTERFACE       ! Uplus(s) is a rectangular matrix
  FUNCTION Uplus(si,ti,numTspins,Tspins,upc,p,q) RESULT  (w)
    INTEGER, INTENT(IN) :: si,ti, numTspins,p, q !spins Tspins(s), Tspins(t)
    REAL, INTENT(IN) :: Tspins(numTspins),upc(0:(numTspins+1),0:(numTspins+1))
    REAL :: sigma,rho                     ! spin components
    REAL :: w(NINT(2*Tspins(si)+1),NINT(2*Tspins(ti)+1))
    INTEGER ::  i,j,sx2,tx2 ! use two times the spins making integers
  END FUNCTION
END INTERFACE
INTERFACE           ! Vplus(s) is a rectangular matrix
  FUNCTION Vplus(si,ti,numTspins,Tspins,upc,p,q) RESULT  (w)
    INTEGER, INTENT(IN) :: si,ti, numTspins,p,q !spins Tspins(s), Tspins(t)
    REAL, INTENT(IN) :: Tspins(numTspins),upc(0:(numTspins+1),0:(numTspins+1))
    REAL :: sigma,rho                  ! spin components
    REAL :: w(NINT(2*Tspins(si)+1),NINT(2*Tspins(ti)+1))
    INTEGER ::  i,j,sx2,tx2  ! use two times the spins making integers
  END FUNCTION
END INTERFACE
!
!------------------------2. Type declarations -----------------------
!
INTEGER ::  p0,q0,p,q          ! integers p0,q0 identify the su(3) irrep
INTEGER ::  numTspins,dimREP    ! the number of T-spins, TUV matrix dimension
INTEGER  ::  s2,i,j,k,m,n,sx2,tx2,q1,IFq,IFj,si,ti    ! dummy indices
REAL, ALLOCATABLE :: zeroMatrix(:,:), unitMatrix(:,:)
REAL, ALLOCATABLE :: Tspins(:),U3firsts(:)
REAL :: sigma,s,t    !   dummy spin variables
REAL, ALLOCATABLE :: TplusMat(:,:),TminusMat(:,:),TthreeMat(:,:),UplusMat(:,:)
REAL, ALLOCATABLE ::   UpMat(:,:), VpMat(:,:)
! The 8 matrices produced by the program:
REAL,ALLOCATABLE :: Tp(:,:),Tm(:,:),T3(:,:),Up(:,:),Um(:,:),U3(:,:)
REAL,ALLOCATABLE :: Vp(:,:),Vm(:,:)
REAL:: MaxErr,MaxErrLimit ! Max error found in required equations, tolerance
REAL, ALLOCATABLE:: upc2(:,:),upc(:,:) ! aux. quantities for Up, Um, Vp, Vm
!
! XYcomm - commutation relation: XYcomm =  [X,Y] - commutator  = 0
REAL,ALLOCATABLE:: T3Tpcomm(:,:),T3Tmcomm(:,:),T3Upcomm(:,:),T3Umcomm(:,:)
REAL,ALLOCATABLE:: T3U3comm(:,:),T3Vpcomm(:,:),T3Vmcomm(:,:),TpTmcomm(:,:)
REAL,ALLOCATABLE:: TpUpcomm(:,:),TpUmcomm(:,:),TpU3comm(:,:),TpVpcomm(:,:)
REAL,ALLOCATABLE:: TpVmcomm(:,:),TmUpcomm(:,:),TmUmcomm(:,:),TmU3comm(:,:)
REAL,ALLOCATABLE:: TmVpcomm(:,:),TmVmcomm(:,:),U3Upcomm(:,:),U3Umcomm(:,:)
REAL,ALLOCATABLE:: U3Vpcomm(:,:),U3Vmcomm(:,:),UpUmcomm(:,:),UpVpcomm(:,:)
REAL,ALLOCATABLE:: UpVmcomm(:,:),UmVpcomm(:,:),UmVmcomm(:,:),VpVmcomm(:,:)
REAL,ALLOCATABLE:: SU3identity(:,:)
LOGICAL:: p0GEq0,noNaNs,MaxErrNotTooBig  ! p0 >= q0?, matrices all numerical?
CHARACTER (len=20) :: file_name         ! file name for output data file
!
!------------------------3. Begin -----------------------
!
! Input
    p0 = 3    ! the program finds matrices for the (p0,q0)-irrep
    q0 = 3
   MaxErrLimit = 1.E-4 ! Largest allowed error in any component of the 29 eqns
!
    dimREP = (p0+1)*(q0+1)*(p0+q0+2)/2 !Matrices are nxn square, n = dimREP
WRITE(*,*)    'The SU(3) irrep for (p0,q0) = ',(/p0,q0/), ' .'
WRITE(*,*)    'The error in any component of the 29 eqns <= ',MaxErrLimit
WRITE(*,*)    'The dimension of the SU(3) irrep is dimREP = ',dimREP, ' .'
WRITE(*,*)
!
!------------------------4. Preliminaries -----------------------
!
  IF (p0 .GE. q0) THEN  ! The formulas are set up for p >= q.
    p0GEq0 = .TRUE.
    p = p0
    q = q0
    ELSE IF (p0 < q0) THEN  !When p < q, we swap p and q for the calculation
        p0GEq0 = .FALSE.    ! and take the negative transpose.
        p = q0
        q = p0
  END IF
!
    ALLOCATE( unitMatrix(dimREP,dimREP)) ! unit matrix, a.k.a. identity matrix
    ALLOCATE( zeroMatrix(dimREP,dimREP))    ! all components vanish
    DO  i = 1,dimREP
        DO  j = 1,dimREP
            unitMatrix(i,j) = 0.  ! the unit matrix has zeros everywhere,except
        END DO
            unitMatrix(i,i) = 1.    ! ones along the diagonal.
    END DO
!
            zeroMatrix = unitMatrix - unitMatrix      ! all components vanish
!
!-----------------5. Make the Tp, Tm, T3, U3 matrices -----------------------
!
    numTspins = (p+1)*(q+1)         ! A formula for the number of Tspins
    ALLOCATE( Tspins(numTspins))
n=1                                 ! Make the list of Tspins
DO s2=1,q
      DO m = 1, s2
         Tspins(n) = Real(s2-1)/2.  !Tp,Tm,T3 satisy the su(2) Lie algebra
         n=n+1            ! Tp,Tm,T3 have blocks of irreps of SU(2) along the
      END DO              ! diagonal.
   END DO              ! Tspins is the list of the SU(2) spins for the blocks.
DO s2=q,p
      DO m = 1, q+1
         Tspins(n) = Real(s2)/2.
         n=n+1
      END DO
   END DO
DO s2=p+1,p+q
      DO m = 1,q+1-s2+p
         Tspins(n) = Real(s2)/2.
         n=n+1
      END DO
   END DO
   WRITE(*,*) 'Check the number of Tspins with the formula: numTspins, &
                SIZE(Tspins) ', numTspins,'?=?', SIZE(Tspins)
!
ALLOCATE( U3firsts(numTspins))  ! Make the list of U3firsts
    n=1
        DO i=1,q
            DO j = 1, i
                    U3firsts(n) =REAL(-(p - q) - 2*(i - 1) + 3*(j - 1))/2.
                    n=n+1           ! Up,Um,U3 also satisfy the su(2) algebra
            END DO              ! and U3 is diagonal.
        END DO                  ! U3 has the same block diagonal structure
        DO i=0,p-q              ! as Tp,Tm,T3
            DO j = 0, q      ! U3firsts are the first components in each block
                    U3firsts(n) = REAL(-p - q + i + 3*j)/2.
                    n=n+1
            END DO
        END DO
        DO i=1,q
            DO j = 1,q-i+1
                    U3firsts(n) = Real(1 - 2*q + (i - 1) + 3*(j - 1))/2.
                    n=n+1
            END DO
        END DO
!
ALLOCATE( Tp(dimREP,dimREP))    ! Make U3, Tp, Tm, T3
ALLOCATE( Tm(dimREP,dimREP))
ALLOCATE( T3(dimREP,dimREP))
ALLOCATE( U3(dimREP,dimREP))
    DO  i = 1,dimREP
        DO  j = 1,dimREP
            Tp(i,j) = 0.   ! Initially, set all components to zero, T(i,j) = 0
            Tm(i,j) = 0.
            T3(i,j) = 0.
            U3(i,j) = 0.    ! and U3(i,j) = 0, initially.
        END DO
    END DO
!
    DO  n = 1,numTspins                 !Tplus(s),Tminus(s),Tthree(s) are the
        DO i=1, NINT(2.*Tspins(n)+1)  !standard SU(2) spin matrices for spin s
U3(i+sumTmatrixDims(n,numTspins,Tspins),i+sumTmatrixDims(n,numTspins,Tspins))&
                                                = U3firsts(n)+0.5*REAL(i-1)
            DO j=1, NINT(2.*Tspins(n)+1)
                TplusMat = Tplus(Tspins(n))
Tp(i+sumTmatrixDims(n,numTspins,Tspins),j+sumTmatrixDims(n,numTspins,Tspins))&
                                                = TplusMat(i,j)
                TminusMat = Tminus(Tspins(n))
Tm(i+sumTmatrixDims(n,numTspins,Tspins),j+sumTmatrixDims(n,numTspins,Tspins))&
                                                = TminusMat(i,j)
                TthreeMat = Tthree(Tspins(n))
T3(i+sumTmatrixDims(n,numTspins,Tspins),j+sumTmatrixDims(n,numTspins,Tspins))&
                                                = TthreeMat(i,j)
            END DO
        END DO
    END DO
!
!--------------------6. Make the Up, Vp, Um, Vm matrices ---------------------
!
!                          Calculate the auxiliary matrices 'upc' and 'upc2',
ALLOCATE( upc2(0:(numTspins+1),  0:(numTspins+1)))  !where upc2 = upc squared
ALLOCATE( upc(0:(numTspins+1),  0:(numTspins+1)))
    DO i = 0,numTspins                              ! Null values initially
        DO  j = 0,numTspins
                upc2(i,j) = 0.
                upc(i,j) = 0.
        END DO
    END DO
    IF  (q==1) THEN                         ! A special case of upc2 and upc
                upc2(3,1) = REAL(p*(q+2))/2.
                upc(3,1) = SQRT(upc2(3,1))
    END IF
    DO q1 = 1,(q-1)          ! uppertopcap, please see text "Formulas for
        DO  i = n0(q1), n0(q1+1)     !SU(3) matrices" for the choice of names
    upc2(i,i+q1) = &
   REAL(n0(q1+1)-i)*REAL(p+(n0(q1+1)-i) +1)*REAL(q-(n0(q1+1)-i) +1)/REAL(q1+1)
                        upc(i,i+q1) = SQRT(upc2(i,i+q1))
        END DO
    END DO
    DO q1 = 1,(q-1)      ! lowertopcap. There is a top cap, the middle section,
        DO  j = n0(q1), n0(q1+1)-1      ! and the bottom cap.
    upc2(j+q1+1,j) = &
    REAL(j-n0(q1)+1)*REAL(p-(j-n0(q1)) )*REAL(q+j-n0(q1) +2)/REAL(q1*(q1+1))
                        upc(j+q1+1,j) = SQRT(upc2(j+q1+1,j))
        END DO
    END DO
    DO i = n0(q), numTspins-q*(q+1)/2          ! upperdiagonal
    upc2(i,i+q) = &
REAL(q-modA(i,q))*REAL(p+q+1-modA(i,q))*REAL(1+modA(i,q))/REAL(q+1+floorA(i,q))
                        upc(i,i+q) = SQRT(upc2(i,i+q))
    END DO
    DO q1 = 1,(q-1)                            ! upperbottomcap
        DO  i = numTspins - n0(q1+1) - q1 + 1, numTspins - n0(q1) - q1 + 1
    upc2(i,i+q1) = &
    REAL(n0(q1+1)+i-numTspins +q1-1)*REAL(p+q-q1-n0(q1+1)+numTspins +3-i)*&
    REAL(q - q1 - n0(q1+1) + numTspins + 2 - i)/REAL(p + q - q1+2)
                        upc(i,i+q1) = SQRT(upc2(i,i+q1) )
        END DO
    END DO
    DO q1 = 1,(q-1)                            ! lowerbottomcap
        DO  j = numTspins - n0(q1+1)+1 - q1,numTspins - n0(q1) - q1
    upc2(j+q1+1,j) = &
    REAL(numTspins - q1 - n0(q1)+1 - j)*REAL(p - numTspins+q1+n0(q1)+j) *&
    REAL(p+q+j - numTspins+q1+n0(q1)+1)/REAL((p+q - q1+1)*(p+q - q1+2))
                        upc(j+q1+1,j) = SQRT(upc2(j+q1+1,j))
        END DO
    END DO
    IF (q == 1)    THEN          ! lowerdiagonal
            IFq = 1                                     !First, set IFq
        ELSE IF (q /= 1)    THEN
            IFq = 0
    END IF                  !The lowerdiagonal is last because it depends
        IF (q >= 1)  THEN       ! on the previous upc2, upc values.
            DO j = n0(q) + IFq, n0(q) + (p - q+2)*(q+1)-3
                IF (j<n0(q+1))    THEN
                        IFj = 0
                ELSE IF (j>=n0(q+1))    THEN
                        IFj = -1
                END IF
    upc2(j+q+1,j) = &
        -1.+upc2(j,j-q+IFj)+upc2(j-q+1+IFj,j)/(2.*Tspins(j))-&
        upc2(j,j+q)/(2.*Tspins(j)+1.)
                        upc(j+q+1,j) = SQRT(upc2(j+q+1,j))
            END DO
        ELSE IF (q < 1) THEN        ! 'If q = 0 ...' is the same because
            DO i = 1,p             ! the only possible value of q<1 is q = 0
                        upc2(i+1,i) = REAL(p-i+1)
                        upc(i+1,i) = SQRT(upc2(i+1,i) )
            END DO
        END IF
!
ALLOCATE( Up(dimREP,dimREP))        ! Up
    DO i = 1,dimREP
        DO  j = 1,dimREP            ! Null values initially
                Up(i,j) = 0.
        END DO
    END DO
                UpMat = zeroMatrix    ! Set UpMat
    DO  si = 1,numTspins        ! Calculate the nonzero components of Up
        DO  ti = 1,numTspins
            DO i= 1,NINT(2.*Tspins(si)+1.)
                DO j= 1,NINT(2.*Tspins(ti)+1)
    IF ((NINT(2.*Tspins(ti))==NINT(2.*Tspins(si)+1.))&  !If(t=s+1/2)&
        .AND.(NINT(U3firsts(ti)-U3firsts(si))==-1)) THEN    !(U3(t)=U3(s)-1...
            UpMat = Uplus(si,ti,numTspins,Tspins,upc,p,q)
Up(i+sumTmatrixDims(si,numTspins,Tspins),j+sumTmatrixDims(ti,numTspins,Tspins))&
            = UpMat(i,j)
            UpMat = zeroMatrix          ! Reset UpMat
    END IF
    IF ((NINT(2.*Tspins(ti))==NINT(2.*Tspins(si)-1.))&  !If(t=s-1/2)&
        .AND.(NINT(U3firsts(si)-U3firsts(ti))==+1)) THEN    !(U3(t)=U3(s)-1...
            UpMat = Uplus(si,ti,numTspins,Tspins,upc,p,q)
Up(i+sumTmatrixDims(si,numTspins,Tspins),j+sumTmatrixDims(ti,numTspins,Tspins))&
            = UpMat(i,j)
            UpMat = zeroMatrix          ! Reset UpMat
                    END IF
                END DO          ! Since spins s and t differ by 1/2,
            END DO              ! the UpMat = Uplus blocks are rectangular
        END DO
 END DO
!
ALLOCATE( Vp(dimREP,dimREP))            ! Vp
    DO i = 1,dimREP
        DO  j = 1,dimREP
                Vp(i,j) = 0.            ! Null values initially
        END DO
    END DO
                VpMat = zeroMatrix    ! Set VpMat
    DO  si = 1,numTspins             ! Calculate the nonzero components of Vp
        DO  ti = 1,numTspins
            DO i=1, NINT(2.*Tspins(si)+1.)
                DO j=1, NINT(2.*Tspins(ti)+1)
    IF ((NINT(2.*Tspins(ti))==NINT(2.*Tspins(si)+1.))&  !If(t=s+1/2)&
        .AND.(NINT(U3firsts(ti)-U3firsts(si))==-1)) THEN    !(U3(t)=U3(s)-1...
        VpMat = Vplus(si,ti,numTspins,Tspins,upc,p,q)
Vp(i+sumTmatrixDims(si,numTspins,Tspins),j+sumTmatrixDims(ti,numTspins,Tspins))&
        =VpMat(i,j)
        VpMat = zeroMatrix       ! Reset VpMat
    END IF
    IF ((NINT(2.*Tspins(ti))==NINT(2.*Tspins(si)-1.))&   !If(t=s-1/2)&
        .AND.(NINT(U3firsts(si)-U3firsts(ti))==+1)) THEN    !(U3(t)=U3(s)-1...
            VpMat = Vplus(si,ti,numTspins,Tspins,upc,p,q)
Vp(i+sumTmatrixDims(si,numTspins,Tspins),j+sumTmatrixDims(ti,numTspins,Tspins))&
        =VpMat(i,j)
        VpMat = zeroMatrix      ! Reset VpMat
                    END IF
                END DO
            END DO
        END DO
    END DO
!
ALLOCATE( Um(dimREP,dimREP))        ! Make the Um and Vm matrices
ALLOCATE( Vm(dimREP,dimREP))
  DO i = 1,dimREP
    DO  j = 1,dimREP
        Um(i,j) = Up(j,i)           ! Transpose Up to get Um
        Vm(i,j) = Vp(j,i)           ! Transpose Vp to get Vm
    END DO
  END DO
! If the given irrep integers p0 and q0 satisfy  p0 >= q0, then these
! 8 matrices make the (p0,q0) irrep.
! If  p0 < q0, then we calculated with p = q0 and q = p0. The negative
! transpose of the 8 matrices of the (p,q)-irrep make the (p0,q0) irrep.
  IF (.NOT.p0GEq0) THEN !When p0<q0,we need one more step to get the matrices.
    Tp = -TRANSPOSE(Tp)
    Tm = -TRANSPOSE(Tm)
    T3 = -TRANSPOSE(T3)
    Up = -TRANSPOSE(Up)
    Um = -TRANSPOSE(Um)
    U3 = -TRANSPOSE(U3)
    Vp = -TRANSPOSE(Vp)
    Vm = -TRANSPOSE(Vm)
  END IF
!
  ! The 8 matrices Tp,Tm,T3,Up,Um,U3,Vp,Vm have been calculated.
!
!------7. Check that all matrix components are numbers, no 'NaN'-------------
    n=0        !Check that all components of the matrices Tp,Tm,T3 are numbers
    DO i=1,dimREP   ! and there are no 'NaN' elements(Not A Number).
        DO j = 1,dimREP
            IF (Tp(i,j) /= Tp(i,j)) THEN                   ! Check Tp
                    WRITE(*,*) 'Suspected NaN: i,j,Tp(i,j)',i,j,Tp(i,j)
                    n=n+1
            END IF
                IF (Tm(i,j) /= Tm(i,j)) THEN            ! Check Tm
                        WRITE(*,*) 'Suspected NaN: i,j,Tm(i,j)',i,j,Tm(i,j)
                        n=n+1
                END IF
            IF (T3(i,j) /= T3(i,j)) THEN                ! Check T3
                    WRITE(*,*) 'Suspected NaN: i,j,T3(i,j)',i,j,T3(i,j)
                    n=n+1
            END IF
        END DO
    END DO
        IF (n==0) THEN
                noNaNs=.TRUE.
!               WRITE(*,*) 'Tp,Tm,T3 are filled with numbers, no"NaN".'
        ELSE IF (n>0) THEN
                noNaNs=.FALSE.
 !              WRITE(*,*) 'Tp,Tm,T3 have at least one nonnumerical element.'
        END IF
    n=0        !Check that all components of the matrices Up,Um,U3 are numbers
        DO i=1,dimREP   ! and there are no 'NaN' elements(Not A Number).
            DO j = 1,dimREP
                IF (Up(i,j) /= Up(i,j)) THEN                ! Check Up
                        WRITE(*,*) 'Suspected NaN: i,j,Up(i,j)',i,j,Up(i,j)
                        n=n+1
                END IF
                    IF (Um(i,j) /= Um(i,j)) THEN             ! Check Um
                        WRITE(*,*) 'Suspected NaN: i,j,Um(i,j)',i,j,Um(i,j)
                            n=n+1
                    END IF
                IF (U3(i,j) /= U3(i,j)) THEN                     ! Check U3
                        WRITE(*,*) 'Suspected NaN: i,j,U3(i,j)',i,j,U3(i,j)
                        n=n+1
                END IF
            END DO
        END DO
            IF (n==0) THEN
                    noNaNs=.TRUE..AND. noNaNs
!                   WRITE(*,*) 'Up,Um,U3 are filled with numbers, no "NaN".'
            ELSE IF (n>0) THEN
                    noNaNs=.FALSE.
 !               WRITE(*,*) 'Up,Um,U3 have at least one nonnumerical element.'
            END IF
    n=0           !Check that all components of the matrices Vp,Vm are numbers
        DO i=1,dimREP   ! and there are no 'NaN' elements(Not A Number).
            DO j = 1,dimREP
                IF (Vp(i,j) /= Vp(i,j)) THEN                ! Check Vp
                        WRITE(*,*) 'Suspected NaN: i,j,Vp(i,j)',i,j,Vp(i,j)
                        n=n+1
                END IF
                    IF (Vm(i,j) /= Vm(i,j)) THEN            ! Check Vm
                            WRITE(*,*) 'Suspected NaN: i,j,Vm(i,j)',i,j,Vm(i,j)
                            n=n+1
                    END IF
            END DO
        END DO
            IF (n==0) THEN
                    noNaNs=.TRUE. .AND. noNaNs
!                   WRITE(*,*) 'Vp,Vm are filled with numbers, no "NaN".'
            ELSE IF (n>0) THEN
                    noNaNs=.FALSE.
  !                 WRITE(*,*) 'Vp,Vm have at least one nonnumerical element.'
            END IF
!
!------8. Check 28 commutator relations and an invariance equation -----------
!
n = dimREP
ALLOCATE(T3Tpcomm(n,n),T3Tmcomm(n,n),T3Upcomm(n,n),T3Umcomm(n,n))
ALLOCATE(T3U3comm(n,n),T3Vpcomm(n,n),T3Vmcomm(n,n),TpTmcomm(n,n))
ALLOCATE(TpUpcomm(n,n),TpUmcomm(n,n),TpU3comm(n,n),TpVpcomm(n,n))
ALLOCATE(TpVmcomm(n,n),TmUpcomm(n,n),TmUmcomm(n,n),TmU3comm(n,n))
ALLOCATE(TmVpcomm(n,n),TmVmcomm(n,n),U3Upcomm(n,n),U3Umcomm(n,n))
ALLOCATE(U3Vpcomm(n,n),U3Vmcomm(n,n),UpUmcomm(n,n),UpVpcomm(n,n))
ALLOCATE(UpVmcomm(n,n),UmVpcomm(n,n),UmVmcomm(n,n),VpVmcomm(n,n))
ALLOCATE(SU3identity(n,n))
        T3Tpcomm = MATMUL(T3,Tp) - MATMUL(Tp,T3) - Tp                   ! 1
        MaxErr = MAXVAL(ABS( T3Tpcomm ))
!   WRITE(*,*)  'MaxErr last comm = ', MAXVAL(ABS( T3Tpcomm )), 1
!   WRITE(*,*)      'Furthest from zero in comm relations ', MaxErr
            T3Tmcomm = MATMUL(T3,Tm) - MATMUL(Tm,T3) + Tm
            MaxErr = MAXVAL(ABS( (/MaxErr, T3Tmcomm/) ))
                T3Upcomm = MATMUL(T3,Up) - MATMUL(Up,T3) + Up/2.
                MaxErr = MAXVAL(ABS( (/MaxErr, T3Upcomm/) ))
                    T3Umcomm = MATMUL(T3,Um) - MATMUL(Um,T3) - Um/2.
                    MaxErr = MAXVAL(ABS( (/MaxErr, T3Umcomm/) ))
                        T3U3comm = MATMUL(T3,U3) - MATMUL(U3,T3)        ! 5
                        MaxErr = MAXVAL(ABS( (/MaxErr, T3U3comm/) ))
                            T3Vpcomm = MATMUL(T3,Vp) - MATMUL(Vp,T3) - Vp/2.
                            MaxErr = MAXVAL(ABS( (/MaxErr, T3Vpcomm/) ))
                              T3Vmcomm = MATMUL(T3,Vm) - MATMUL(Vm,T3) + Vm/2.
                                MaxErr = MAXVAL(ABS( (/MaxErr, T3Vmcomm/) ))
        TpTmcomm = MATMUL(Tp,Tm) - MATMUL(Tm,Tp) - 2.*T3
        MaxErr = MAXVAL(ABS( (/MaxErr, TpTmcomm/) ))
            TpUpcomm = MATMUL(Tp,Up) - MATMUL(Up,Tp) - Vp
            MaxErr = MAXVAL(ABS( (/MaxErr, TpUpcomm/) ))
                TpUmcomm = MATMUL(Tp,Um) - MATMUL(Um,Tp)                ! 10
                MaxErr = MAXVAL(ABS( (/MaxErr, TpUmcomm/) ))
                    TpU3comm = MATMUL(Tp,U3) - MATMUL(U3,Tp) - Tp/2.
                    MaxErr = MAXVAL(ABS( (/MaxErr, TpU3comm/) ))
                        TpVpcomm = MATMUL(Tp,Vp) - MATMUL(Vp,Tp)
                        MaxErr = MAXVAL(ABS( (/MaxErr, TpVpcomm/) ))
                            TpVmcomm = MATMUL(Tp,Vm) - MATMUL(Vm,Tp) + Um
                            MaxErr = MAXVAL(ABS( (/MaxErr, TpVmcomm/) ))
        TmUpcomm = MATMUL(Tm,Up) - MATMUL(Up,Tm)
        MaxErr = MAXVAL(ABS( (/MaxErr, TmUpcomm/) ))
            TmUmcomm = MATMUL(Tm,Um) - MATMUL(Um,Tm) + Vm               ! 15
            MaxErr = MAXVAL(ABS( (/MaxErr, TmUmcomm/) ))
                TmU3comm = MATMUL(Tm,U3) - MATMUL(U3,Tm) + Tm/2.
                MaxErr = MAXVAL(ABS( (/MaxErr, TmU3comm/) ))
                    TmVpcomm = MATMUL(Tm,Vp) - MATMUL(Vp,Tm) - Up
                    MaxErr = MAXVAL(ABS( (/MaxErr, TmVpcomm/) ))
                        TmVmcomm = MATMUL(Tm,Vm) - MATMUL(Vm,Tm)
                        MaxErr = MAXVAL(ABS( (/MaxErr, TmVmcomm/) ))
        U3Upcomm = MATMUL(U3,Up) - MATMUL(Up,U3) - Up
        MaxErr = MAXVAL(ABS( (/MaxErr, U3Upcomm/) ))
            U3Umcomm = MATMUL(U3,Um) - MATMUL(Um,U3) + Um               ! 20
            MaxErr = MAXVAL(ABS( (/MaxErr, U3Umcomm/) ))
                U3Vpcomm = MATMUL(U3,Vp) - MATMUL(Vp,U3) - Vp/2.
                MaxErr = MAXVAL(ABS( (/MaxErr, U3Vpcomm/) ))
                    U3Vmcomm = MATMUL(U3,Vm) - MATMUL(Vm,U3) + Vm/2.
                    MaxErr = MAXVAL(ABS( (/MaxErr, U3Vmcomm/) ))
        UpUmcomm = MATMUL(Up,Um) - MATMUL(Um,Up) - 2.*U3
        MaxErr = MAXVAL(ABS( (/MaxErr, UpUmcomm/) ))
            UpVpcomm = MATMUL(Up,Vp) - MATMUL(Vp,Up)
            MaxErr = MAXVAL(ABS( (/MaxErr, UpVpcomm/) ))
                UpVmcomm = MATMUL(Up,Vm) - MATMUL(Vm,Up) - Tm           ! 25
                MaxErr = MAXVAL(ABS( (/MaxErr, UpVmcomm/) ))
        UmVpcomm = MATMUL(Um,Vp) - MATMUL(Vp,Um) + Tp
        MaxErr = MAXVAL(ABS( (/MaxErr, UmVpcomm/) ))
            UmVmcomm = MATMUL(Um,Vm) - MATMUL(Vm,Um)
            MaxErr = MAXVAL(ABS( (/MaxErr, UmVmcomm/) ))
        VpVmcomm = MATMUL(Vp,Vm) - MATMUL(Vm,Vp) - 2.*U3 - 2.*T3        ! 28
        MaxErr = MAXVAL(ABS( (/MaxErr, VpVmcomm/) ))
!
        SU3identity = (MATMUL(Tp,Tm)+MATMUL(Tm,Tp)+MATMUL(Up,Um)+ &
            MATMUL(Um,Up)+MATMUL(Vp,Vm)+MATMUL(Vm,Vp))/2.+MATMUL(T3,T3)+ &
            MATMUL(2.*U3+T3,2.*U3+T3)/3.-(REAL(p**2+p*q+q**2)/3.+ &
            REAL(p + q))*unitMatrix                                     ! 29
        MaxErr = MAXVAL(ABS( (/MaxErr, SU3identity/) ))
!
    WRITE(*,*)  'The 28 commutation relations plus one quadratic invariance &
                expression are set up to have null components.'
WRITE(*,*) 'Substituting the matrices in the eqns gives a largest error of '
    WRITE(*,*)  '                  ', MaxErr, ' , '
    WRITE(*,*)  'which is the largest error in any matrix component of the 29 &
                   expressions.'
  WRITE(*,*)'(The allowed tolerance is set at MaxErrLimit = ',MaxErrLimit,'.)'
    WRITE(*,*)
!
    IF(MaxErr.LE.MaxErrLimit) THEN    ! largest error within tolerance?
        MaxErrNotTooBig = .TRUE.        ! TRUE means Yes.
    ELSE IF (MaxErr >MaxErrLimit)    THEN
        MaxErrNotTooBig = .FALSE.       ! FALSE means No, not within tolerance
    END IF
!
!----------9. Save the results to a file, end program -----------------------
!
    WRITE(*,*)"The matrix elements must be numbers and the 29 relations' max &
                error must be within set tolerance."
    WRITE(*,*) 'Both must be TRUE to get an output file:'
    WRITE(*,*) '   noNaNs,  MaxErrNotTooBig :   ', noNaNs,MaxErrNotTooBig,' .'
 !
 !MaxErrNotTooBig = .FALSE.    ! Use this statement to test the failure option
 !noNaNs = .FALSE.             ! Use this statement to test the failure option
    IF ((.NOT.noNaNs).OR.(.NOT.MaxErrNotTooBig)) THEN       ! Failure(s) found
        WRITE (file_name,"('p',i0,'q',i0,'SU3FAILS.dat')")p0,q0
        OPEN(Unit=5,file=file_name) ! The failure occurs with p0 and q0
        WRITE(5,*)  'The program failed for irrep p0, q0 = ', p0,q0
        CLOSE(5)
    END IF
!
    IF ((noNaNs).AND.(MaxErrNotTooBig)) THEN             ! No Failure(s) found
        WRITE (file_name,"('p',i0,'q',i0,'SU3TUV.dat')")p0,q0
        OPEN(Unit=5,file=file_name)
        WRITE(5,3000) p0,q0, MaxErr   ! Start the data file with p0,q0,MaxErr.
        CLOSE(5)    ! Next, the 8*dimREP**2 components of the 8 basis matrices
    OPEN(Unit=5,file=file_name,STATUS='OLD', POSITION='APPEND') ! Append T,U,V
    WRITE(5,4000) TRANSPOSE(Tp), TRANSPOSE(Tm), TRANSPOSE(T3),TRANSPOSE(Up), &
                    TRANSPOSE(Um), TRANSPOSE(U3), TRANSPOSE(Vp), TRANSPOSE(Vm)
      CLOSE(5)
    END IF
!
  3000 FORMAT(2I3,1E16.7) !FORMAT statement for two integers and a real number
4000 FORMAT(5E16.7,/)!The FORMAT statement for the matrices,5 numbers per line
!
        END PROGRAM Su3Formulas
!
!-----------------10. External functions, end-of-file-----------------------
!
FUNCTION n0(q) RESULT  (w)    ! n0 = 1 + q(q-1)/2
    IMPLICIT NONE
    INTEGER :: w     ! n0 - often used in dummy indice limits.
    INTEGER, INTENT(IN) :: q    ! q characterizes the (p,q)-irrep
            w = 1+q*(q-1)/2
END FUNCTION
FUNCTION modA(n,q) RESULT  (w)      ! modA(n,q) = MOD(n-n0(q),q+1)
    IMPLICIT NONE
    INTEGER :: w,n0        ! modA(n,q) - often used in dummy indice limits.
    INTEGER, INTENT(IN) :: n,q  ! irrep is (p,q), n >= n0(q)
            w = MOD(n-n0(q),q+1)
END FUNCTION
FUNCTION floorA(n,q) RESULT  (w)  ! floorA(n,q) = FLOOR( (n-n0(q))/(q+1) )
    IMPLICIT NONE
    INTEGER :: w,n0       ! floor(n,q) - often used in dummy indice limits.
    INTEGER, INTENT(IN) :: n,q  ! irrep is (p,q), n >= n0(q)
            w = FLOOR(Real((n-n0(q)))/(q+1))
END FUNCTION
FUNCTION rp(s,sigma) RESULT  (w)  !
    IMPLICIT NONE        ! standard spin formula,
    REAL :: w               ! S. Weinberg, Ref. [4], Eq. (5.6.17).
    REAL, INTENT(IN) :: s,sigma  ! spin, spin component
            w = SQRT((s-sigma)*(s+sigma+1))
END FUNCTION
FUNCTION rm(s,sigma) RESULT  (w)  !
    IMPLICIT NONE        ! standard spin formula,
    REAL :: w               ! S. Weinberg, Ref. [4], Eq. (5.6.17).
    REAL, INTENT(IN) :: s,sigma  ! spin, spin component
            w = SQRT((s+sigma)*(s-sigma+1))
END FUNCTION
FUNCTION Tplus(s) RESULT  (w)
    IMPLICIT NONE   !Tplus(s) standard spin step up matrix, S. Weinberg, [4]
    REAL, INTENT(IN) :: s   ! the spin
    REAL :: sigma,sigma1,rp     ! spin components, spin function rp(s,sigma)
    REAL :: w(NINT(2*s+1),NINT(2*s+1))
    INTEGER :: i,j,sx2 ! two times the spin is an integer
                sx2=NINT(2.*s)
  DO i=sx2,-sx2,-2    ! i=2*spin component
      DO j = sx2,-sx2,-2
                sigma = REAL(i)/2.   ! spin component
                sigma1 = REAL(j)/2.  ! spin component
        IF (i == j+2)  THEN ! note that (sx2+2-i)/2 = 1,2,3...2s+1, ditto j
                w((sx2+2-i)/2,(sx2+2-j)/2) = rp(s,sigma1)
            ELSE IF (i /= j+2) THEN
                w((sx2+2-i)/2,(sx2+2-j)/2) = 0.
        END IF
      END DO
   END DO
END FUNCTION
FUNCTION Tminus(s) RESULT  (w)
    IMPLICIT NONE  !Tminus(s) standard spin step down matrix, S. Weinberg, [4]
    REAL, INTENT(IN) :: s   ! the spin
        REAL :: sigma,sigma1,rm     ! spin components, spin function rm(s,sigma)
    REAL :: w(NINT(2*s+1),NINT(2*s+1))
    INTEGER ::  i,j,sx2  ! two times the spin is an integer
                sx2=NINT(2.*s)      ! twice the spin
DO i=sx2,-sx2,-2    ! i=2*spin component
      DO j = sx2,-sx2,-2
                sigma = REAL(i)/2.   ! spin component
                sigma1 = REAL(j)/2.  ! spin component
        IF (i == j-2)  THEN    ! note that (sx2+2-i)/2 = 1,2,3...2s+1, ditto j
                    w((sx2+2-i)/2,(sx2+2-j)/2) = rm(s,sigma1)
            ELSE IF (i /= j-2) THEN
                    w((sx2+2-i)/2,(sx2+2-j)/2) = 0.
        END IF
      END DO
   END DO
END FUNCTION
FUNCTION Tthree(s) RESULT  (w)  !Tthree(s), standard spin matrix,
    IMPLICIT NONE               !  S. Weinberg, Ref. [4], Eq. (5.6.16)
    REAL, INTENT(IN) :: s   ! the spin
        REAL :: sigma,sigma1
    REAL :: w(NINT(2*s+1),NINT(2*s+1))
    INTEGER :: i,j,sx2              ! two times the spin is an integer
                sx2=NINT(2.*s)               ! twice the spin
DO i=sx2,-sx2,-2    ! i=2*spin component
      DO j = sx2,-sx2,-2
                sigma = REAL(i)/2.   ! spin component
                sigma1 = REAL(j)/2.  ! spin component
          IF (i == j)  THEN         !  diagonal
                    w((sx2+2-i)/2,(sx2+2-j)/2) = sigma
          ELSE IF (i /= j) THEN !note that (sx2+2-i)/2 = 1,2,3...2s+1, ditto j
                        w((sx2+2-i)/2,(sx2+2-j)/2) = 0.
          END IF
      END DO
   END DO
END FUNCTION
FUNCTION sumTmatrixDims(n,numTspins,Tspins) RESULT  (w)
  IMPLICIT NONE !The nth block starts at the end of the first n-1 Tspin blocks
    INTEGER, INTENT(IN) :: n,numTspins   !  nth spin, number of Tspins
    REAL, INTENT(IN) :: Tspins(numTspins)   ! the list of Tspins
    INTEGER :: i,k,w        ! i,k - dummy indices, the result w is an integer
    REAL :: TspinN            ! dummy spin
                k = 0            ! initialize k
    DO i=1,n-1            ! sum the dimensions of the first n-1 Tspin matrices
                TspinN = Tspins(i)
                k = k + NINT(2.*TspinN)+1  ! add the number of spin components
                                            !for each previous Tspin block
    END DO
                w=k          ! the result is the sum of Tspin block dimensions
END FUNCTION
FUNCTION Uplus(si,ti,numTspins,Tspins,upc,p,q) RESULT  (w)  ! Uplus(s)
  IMPLICIT NONE !Tspins(s),Tspins(t) are spins,one integral, one half integral
    INTEGER, INTENT(IN) :: si,ti,numTspins,p,q !spins Tspins(s), Tspins(t)
    REAL,INTENT(IN):: Tspins(numTspins),upc(0:(numTspins+1),0:(numTspins+1))
    REAL :: sigma,rho                               ! spin components
    REAL :: w(NINT(2*Tspins(si)+1),NINT(2*Tspins(ti)+1))
    INTEGER ::  i,j,sx2, tx2 ! dummys       ! Uplus(s) is a rectangular matrix
                sx2=NINT(2.*Tspins(si))        ! twice the spin is an integer
               tx2=NINT(2.*Tspins(ti)) !we use twice the spin, for convenience
    DO i=sx2,-sx2,-2                ! i=2*spin component sigma
        DO j = tx2,-tx2,-2          ! j=2*spin component rho
                sigma = REAL(i)/2.      ! s-spin component is sigma
                rho = REAL(j)/2.        ! t-spin component is rho
            IF (i == j-1)  THEN     !  i.e., does sigma = rho - 1/2 ?
                IF (Tspins(si)>Tspins(ti)) THEN     ! If s > t ...
                w((sx2+2-i)/2,(tx2+2-j)/2) = upc(si,ti)*SQRT(Tspins(si)-sigma)
                    ELSE IF (Tspins(si)<Tspins(ti)) THEN  ! If s > t ...
                w((sx2+2-i)/2,(tx2+2-j)/2) = &
                    upc(si,ti)*SQRT((1.+Tspins(si)+sigma)/(1.+2.*Tspins(si)))
                END IF ! note that (sx2+2-i)/2 = 1,2,3...2s+1, ditto tx2 and j
            END IF
        END DO
    END DO
END FUNCTION
FUNCTION Vplus(si,ti,numTspins,Tspins,upc,p,q) RESULT  (w)   ! Vplus(s)
  IMPLICIT NONE !Tspins(s),Tspins(t) are spins,one integral, one half integral
    INTEGER, INTENT(IN)  :: si,ti,numTspins,p,q !spins Tspins(s), Tspins(t)
    REAL,INTENT(IN):: Tspins(numTspins),upc(0:(numTspins+1),0:(numTspins+1))
    REAL :: sigma,rho                           ! spin components
    REAL :: w(NINT(2*Tspins(si)+1),NINT(2*Tspins(ti)+1))
    INTEGER ::  i,j, sx2, tx2       ! Vplus(s) is a rectangular matrix
                sx2=NINT(2.*Tspins(si)) ! twice the spin is an integer
               tx2=NINT(2.*Tspins(ti)) !we use twice the spin, for convenience
    DO i=sx2,-sx2,-2                    ! i=2*spin component sigma
        DO j = tx2,-tx2,-2              ! j=2*spin component rho
                    sigma = REAL(i)/2.      ! s-spin component is sigma
                    rho = REAL(j)/2.        ! t-spin component is rho
            IF (i == j+1)  THEN         !  i.e., does sigma = rho + 1/2 ?
                IF (Tspins(si)>Tspins(ti)) THEN    ! If s > t ...
                w((sx2+2-i)/2,(tx2+2-j)/2) = upc(si,ti)*SQRT(Tspins(si)+sigma)
                    ELSE IF (Tspins(si)<Tspins(ti)) THEN  ! If s > t ...
                w((sx2+2-i)/2,(tx2+2-j)/2) = &
                    -upc(si,ti)*SQRT((1.+Tspins(si)-sigma)/(1.+2.*Tspins(si)))
               END IF ! note that (sx2+2-i)/2 = 1,2,3...2s+1,  ditto tx2 and j
            END IF
        END DO
    END DO
END FUNCTION
! end-of-file
\end{verbatim}


\appendix


\section{Exercises and Problems} \label{Pb}

\vspace{0.2cm}
\noindent 1. Rewrite the commutation relations (\ref{commRs2}) - (\ref{commRs5}) with $T^{c}$ = $2T^{3}$ and $U^{c}$ = $2U^{3}.$
 See M. F. O'Reilly, Ref. \cite{GandGandM}.
 
\noindent No answer provided.

\vspace{0.2cm}
\noindent 2. A numerical catalog of matrices for various $(p,q)$ irreps gives the following table of values for the square of one of the $U^{3}$ block unknowns  ${u^{+}_{(ij)}}^2$. Here, $i$ and $j$ identify T spins Eq. (\ref{Ts1}); they are not component indices. Find a formula in terms of $p$ and $q$ for the unknown ${u^{+}_{(ij)}}^2.$ Which unknown is it, i.e. what are the $i$ and $j$ T spin IDs?

\begin{center}
    \begin{tabular}{ | l | c | c | c | c | c | c | }
    \hline
    $(p,q)$ & (1,1) & (2,1) & (3,1) & (4,1) & (5,1) & \\ \hline
    ${u^{+}_{(ij)}}^2$  & 3/2 & 3 & 9/2 & 6 & 15/2 & \\ \hline \hline
    $(p,q)$ & & (2,2) & (3,2) & (4,2) & (5,2) & (6,2)\\ \hline
    ${u^{+}_{(ij)}}^2$ & & 4 & 6 & 8  & 10 & 12 \\  \hline \hline
    $(p,q)$ & & & (3,3) & (4,3) & (5,3) &   \\ \hline
    ${u^{+}_{(ij)}}^2$ & & & 15/2 & 10 & 25/2  &   \\  \hline \hline
    $(p,q)$ & & &  & (4,4) & (5,4) &   \\ \hline
    ${u^{+}_{(ij)}}^2$ & & &  & 12 & 15  &   \\  \hline

    \end{tabular}
\end{center}
 
 \noindent Answer: Formula: ${u^{+}_{(ij)}}^2$ = $p(q+2)/2.$  Unknown T spin block: $(i,j)$ = $(3,1).$

\vspace{0.2cm}
\noindent 3. (a) From Sec. \ref{ups}, determine the ${u^{+}_{(ij)}}^2$ for the $(2,0)$ irrep. Then find the eight matrices $\{T^{\pm},T^{3},U^{\pm},U^{3},V^{\pm} \}.$ (b) Repeat for the $(1,1)$  irrep.

\noindent No answer provided.

\vspace{0.2cm}
\noindent 4. Refer to Fig. 1. Find the $\{T^{3},Y\}$ coordinates (a) of the state with T-spin $s$ = 0; (b) of the two sets of two states each with T-spin $s$ = 1/2; (c) of the nine states with T-spin \linebreak $s$ = 4;  (d) of the two sets of eight states each with T-spin $s$ = 7/2.

\noindent Answer: (a) $\{T^{3},Y\}$ = $\{0,-4/3\}.$ (b) $\{T^{3},Y\}$ = $\{+ 1/2,-1/3\},$ $\{- 1/2,-1/3\}.$ and $\{T^{3},Y\}$ = $\{+ 1/2,-7/3\},$ $\{- 1/2,-7/3\}.$ 
(c) $\{T^{3},Y\}$ = $\{-4,+ 2/3\}$ to $\{+4,+ 2/3\},$ $T^{3}$s separated by $+1/2$ and $Y$ constant at $Y$ = $+2/3.$
(d) $\{T^{3},Y\}$ = $\{-7/2,+ 5/3\}$ to $\{+7/2,+ 5/3\}$ and $\{T^{3},Y\}$ = $\{-7/2,- 1/3\}$ to $\{+7/2,- 1/3\}.$

\vspace{0.2cm}
\noindent 5. To obtain the ${u^{+}_{(ij)}}$ for the $(5,3)$ irrep, the positive square roots of ${u^{+}_{(ij)}}^2$ were chosen. More generally, some or all of these square roots could have been negative. Consider all sets of positive and negative roots. Would all choices of the $\pm$ signs give matrices that satisfy the commutation relations?

\noindent No answer provided.

\vspace{0.2cm}
\noindent 6. Solve the commutation relations for the top cap and upper diagonal of the middle. Do the results agree with those displayed in Sec. \ref{ups}?

\noindent No answer provided.

\section{Figures} \label{Fig}

\begin{figure}[ht] \label{53states}	
\centering
\vspace{0cm}
\hspace{0in}\includegraphics[0,0][300,300]{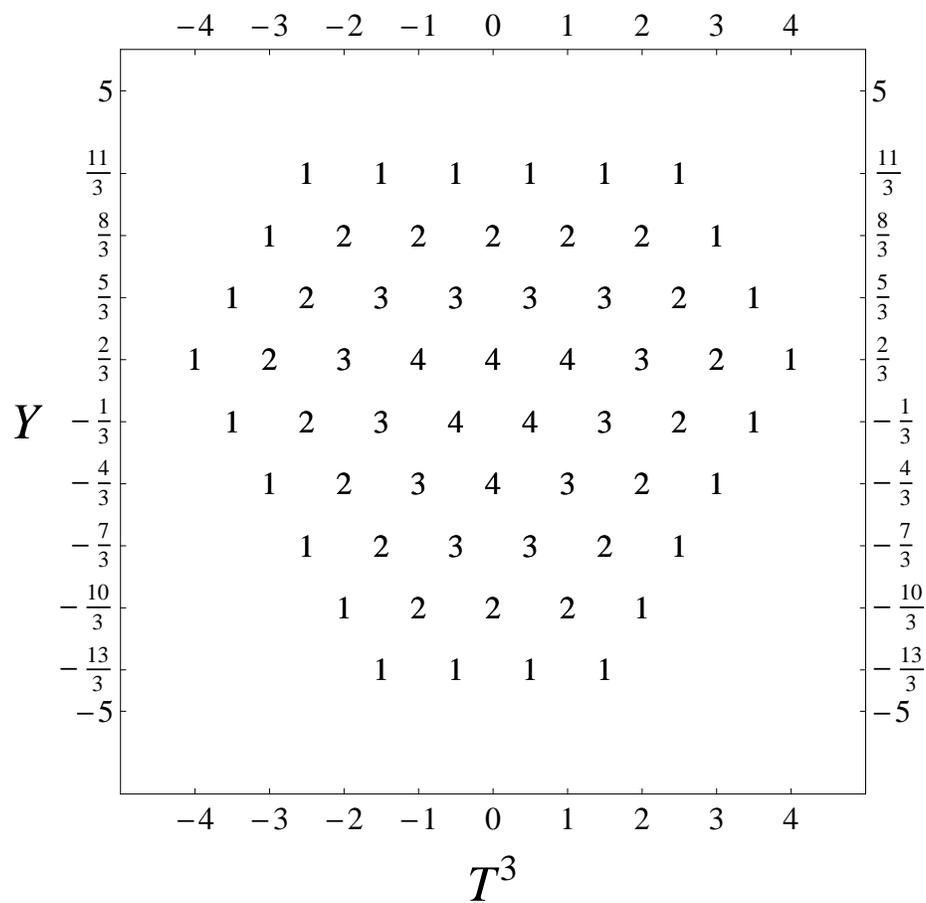}
\caption{{\it{Multiplicity of States for the $(5,3)$ Irrep.}} The number of states with quantum numbers $\{T^{3},Y\}$ where $T^{3}$ is the `3-component' of T-spin   and the `hypercharge' $Y$ is given by $Y$ = $(4/3)U^{3} + (2/3)T^{3},$ with $U^{3}$ is the `3-component' of U-spin. There are $d$ = 120 states total.  }
\end{figure}

\begin{figure}[ht] \label{upc2}	
\centering
\vspace{5cm}
\hspace{0in}\includegraphics[0,0][300,300]{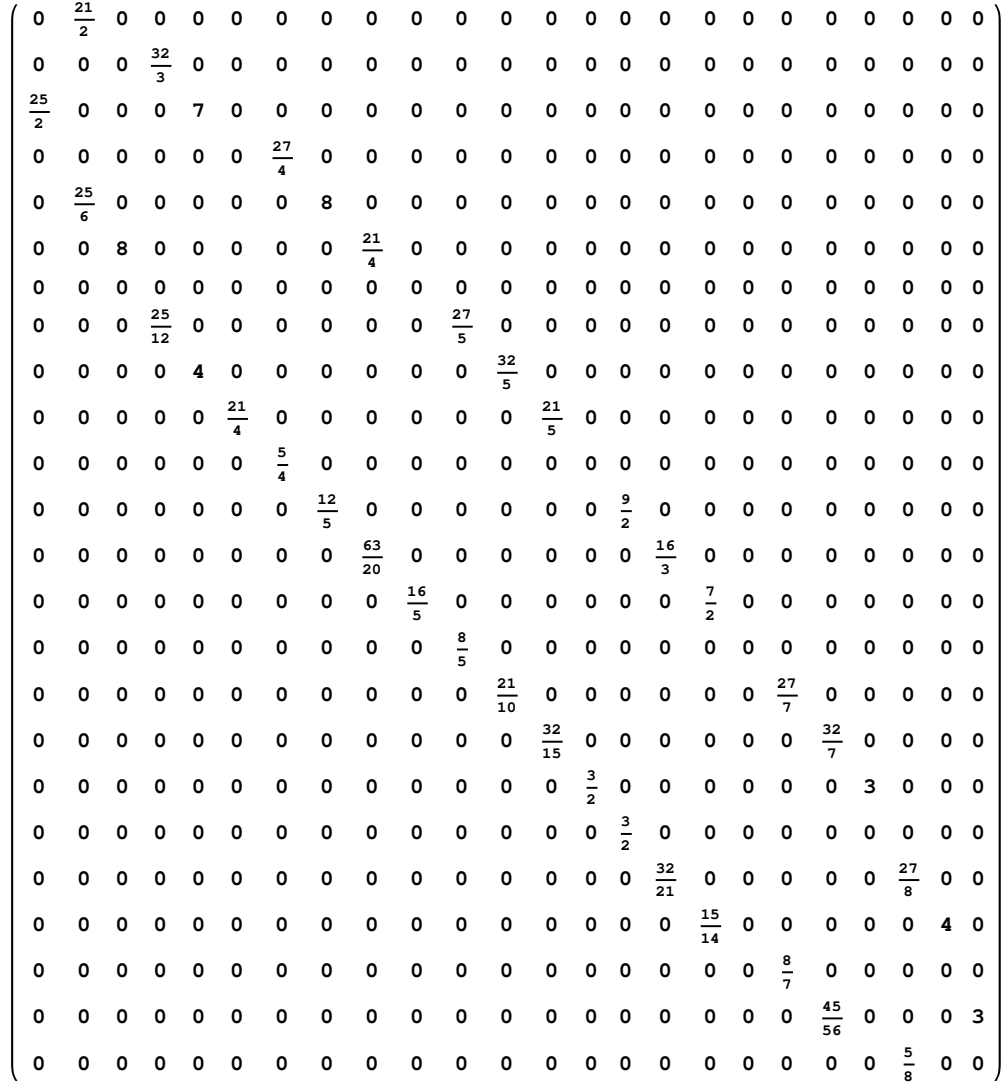}
\caption{{\it{The Matrix of Block Unknowns ${u^{+}_{(ij)}}^2$ for the $(5,3)$ Irrep. }} The components of the $d \times d$ basis matrices are sorted into `blocks' by T-spin, where T-spins label SU(2) irreps. For the $(5,3)$ irrep, there are 24 T-spins, including repeats, making a $24 \times 24$ array of blocks. Recursions and other constraints reduce the number of unknowns from $d^2$ = $120^{2}$ to about 40 for the $(5,3)$ irrep shown here. Note that the nonzero ${u^{+}_{(ij)}}^2$s form upper and lower diagonals in the middle region, enclosed by top and bottom `caps' with staggered upper and lower diagonals of their own. }
\end{figure}

\end{document}